\documentclass[prd,showpacs,nofootinbib,preprintnumbers]{revtex4}

\usepackage{latexsym}
\usepackage{amssymb}
\usepackage{amsmath}
\usepackage{bm}
\usepackage{subfigure}
\usepackage{float}
\usepackage{indentfirst}
\usepackage{xspace}

\usepackage[dvips]{epsfig}
\usepackage{psfrag}
\usepackage{amscd}
\usepackage[matrix,arrow,curve]{xy}
\usepackage{slashbox}

\newcommand{\comment}[1]{}

\newcommand{\hsp}{\hspace{0.15em}}
\newcommand{\nhsp}{\hspace{-0.15em}}
\newcommand{\smhsp}{\hspace{0.06em}}

\newcommand{\dDx}{d^{\hsp D}\nhsp x}

\newcommand{\ds}{\displaystyle}
\newcommand{\Eps}{\mathcal{E}}
\newcommand{\lbr}{\left(}
\newcommand{\rbr}{\right)}

\newcommand{\la}{\lambda}

\newcommand{\om}{\omega}
\newcommand{\al}{\alpha}

\newcommand{\vak}{\varkappa_D}
\newcommand{\pa}{{\partial}}
\newcommand{\ga}{\gamma}
\newcommand{\si}{\sigma}
\newcommand{\Si}{\Sigma}

\newcommand{\fr}{\frac}

\newcommand{\bw}{\begin{widetext}}
\newcommand{\ew}{\end{widetext}}
\newcommand{\be}{\begin{align}}
\newcommand{\ee}{\end{align}}
\newcommand{\ba}{\begin{eqnarray}}
\newcommand{\ea}{\end{eqnarray}}
\newcommand{\non}{\nonumber}

\newcommand{\ep}{\epsilon}

\def\cV{{\cal V}}

\def\r{{\bf r}}

\def\k{{\bf k}}

\def\e{{\rm e}}
\def\RS{\mbox{RS\hsp{}II}\xspace}

\newcommand{\zt}{\dot{z}}

\newcommand{\sgn}{{\rm sgn\smhsp}}

\newcommand{\ah}{ \mathrm{a}}
\newcommand{\bh}{ \mathrm{b}}

\newcommand{\smph}{\vphantom{ d^0_0}}
\newcommand{\vp}{\vphantom{\frac{a}{a}}}

\renewcommand{\vec}{\mathbf}

\renewcommand{\ge}{\geqslant}
\renewcommand{\leq}{\leqslant}
\renewcommand{\geq}{\geqslant}

 \numberwithin{equation}{section}

\begin{document}

\title{Perforation of domain wall by point mass}
\author{D.\,V.\,Gal'tsov\footnote{E-mail: galtsov@phys.msu.ru },
 E.\,Yu.\,Melkumova\footnote{E-mail: elenamelk@mail.ru }, and
P.\,Spirin\footnote{E-mail:  pspirin@physics.uoc.gr}}
\address{ Department of Theoretical Physics, Moscow State University,119899,
Moscow, Russia}  


\begin{abstract}
We investigate collision of a  point particle and an infinitely thin
planar domain wall interacting gravitationally within the linearized
gravity in Minkowski space-time of arbitrary dimension. In this
setting we are able to describe analytically the perforation of the
wall by an impinging particle, showing that it is accompanied by
excitation of the spherical shock branon wave propagating outwards
with the speed of light. Formally, the shock wave is a free solution
of the branon wave equation which has to be added to ensure the
validity of the retarded solution at the perforation point.
Physically, the domain wall gets excited due to the shake caused by
an instantaneous change of sign of the repulsive gravitational
force. This effect is shown to hold, in particular, in four
space-time dimensions, being applicable to the problem of
cosmological domain walls.

\end{abstract}
\maketitle


\section{Introduction}
Gravitational interaction of relativistic extended objects has some
unusual features. It is an essentially relativistic problem even if
their relative velocity  is small, since the brane tension, causing
gravitational repulsion, contributes to interaction on equal footing
with the energy density. The net effect of gravitational interaction
of two branes therefore varies with dimensionality of the
world-volumes and codimension of their embedding into space-time. It
is repulsive for codimension one (domain walls), locally  vanishes
for codimension two (strings) and attractive in other cases.

Another new feature due to the extended nature of branes is
possibility of their free oscillations which may accompany the
generic collision process. While two point particles under collision
just change their momenta, but remain in the same intrinsic state,
the branes will get excited and will not remain in the initial state
even asymptotically. Physically, the most interesting case is
interaction of domain walls with particles in four dimensions. This
was investigated long ago in connection with topological defects in
cosmology. The solution of linearized Einstein equations for
gravitational field of thin planar domain wall was found by Vilenkin
\cite{linear, linear1}. Gravity of domain walls is repulsive, so a
particle impinging on the wall may be reflected at some finite
distance, transferring the momentum to the wall. This gives rise to
the friction force acting upon the wall moving in cosmic plasma
\cite{Vilsh}. Here we will be interested in more subtle effect of
excitation of the wall during the collision, especially in the
situation when the particle energy is enough to reach the wall and
to pierce it.

Fortunately, such  piercing collision may be  treated analytically
with the linearized gravity theory. In fact, one important feature
associated with codimension of  embedding of an extended object into
the bulk is the degree of singularity of the linearized
gravitational field  at its location. Gravitational field diverges
as a negative power of the distance for codimension greater than
two, it diverges logarithmically  for codimension two, but it
remains finite in the case of the domain wall. Therefore, though
generically gravitational collision of two infinitely thin branes is
a singular problem, the  collision  particle -- domain wall turns
out to be tractable within the linearized gravity.

Since gravity of the domain wall remains finite when the particle
pierces it, the  energy-momentum is not exchanged at perforation, so
such particles will not contribute to the friction force. We will
see, however, that the wall does not remain insensitive to piercing
by a point particle, but gets excited. The excitation has a from of
the spherical branon wave arising at the moment of perforation and
propagating outwards with the velocity of light. This effect was
found in five dimensions in the Randall-Sundrum (RS) type of setting
\cite{GMZ} and here it will be shown to exist in arbitrary
space-time dimensions, including the case of domain walls in the
four-dimensional cosmology.

Our treatment of perforation is essentially local, so  in order  to
establish its validity region we should invoke some results obtained
in the full non-linear theory. It is well-known that the linear
approximation for gravity of the domain wall breaks down at large
distances. It is also known \cite{time,Dokh} that no static solution
of full non-linear Einstein equations exists which could fit to the
linearized solution of \cite{linear, linear1}, the consistent
non-linear ansatz being time-dependent. In particular, an exact
solution found by Vilenkin \cite{time} turns out to be a segment of
an accelerated spherical domain wall \cite{IS, IS1} which comes in
from infinity, turns around, and heads back out to infinity.
Subsequently, Linet have shown that the static solution still
 does  exist \cite{linet}, but at the expense of introducing a
cosmological constant in the bulk. More recently domain walls
attracted much attention in the brane-world scenarios \cite{defect,
defect1, defect2, defect3, defect4} and especially in the
Randall-Sundrum models \cite{RS, RS1, RS2}. In fact, the \RS model
with one brane is based on the exact static solution of Einstein
equations with negative cosmological constant. As it could be
expected, this solution reduces to static solution of
five-dimensional linearized Einstein equations in the vicinity of
the brane (after a suitable coordinate transformation \cite{GMZ}),
while corrections due to the cosmological constant enter only in the
second post-linear order. The static domain wall solutions (thick
walls) exist also is field-theoretical models with scalar fields
whose vacuum manifold contains disconnected components, in
particular, supergravities/effective string theories
\cite{Duff:1994an,cvet}.

Apart from calculation of the friction force, gravitational
interaction of domain walls and  more general   $p\hsp -$branes
with massive bodies was studied previously in  other physical
contexts. Some important applications were related to black holes
in the brane-world scenarios (for a recent review and further
references see \cite{Tanahashi:2011xx}). In the Rundall-Sundrum
\cite{RS,RS1,RS2} setup the particles which are allowed to live in
the bulk are expelled from the brane (domain wall) and move along
the geodesics into the AdS bulk \cite{geod, geod2, geod3}. Their
gravity and the corresponding perturbation of the induced metric
on the brane was studied in \cite{Gregory:2000rh}. Black holes,
which in such scenarios  can be created in high energy particle
collisions, also may escape from the brane into the bulk
\cite{escape, escape2, escape3}. The brane -- black-hole system,
including the process of the  merging, was investigated in detail
using the model of a test brane in the black hole background
\cite{BBHS, BBHS2, BBHS3, BBHS4, BBHS5}. Perforation of domain
walls by black holes was qualitatively studied within the
field-theoretical model of hybrid defects (axion)
\cite{ChamEard,Stojkovic:2005zh} suggesting this mechanism as
relevant to the cosmological domain wall problem. Numerical
studies of interaction of black holes with field-theoretical
domain walls are also available \cite{Flachi:2007ev}.

Our approach in this paper is much simpler and it can be considered
complementary to the above studies. We treat both the domain wall
and the bulk particle as test bodies propagating in $D-$dimensional
Minkowski space-time and interacting via linearized gravity. Such a
setting seems adequate to give a local description of the process of
perforation with possibility to treat the domain wall dynamically.

The  paper is organized as follows. In Sec. 2 we derive the brane
metric in the linearized gravity, consider its relation to some
exact solutions and explore interaction of static branes of
different dimensionality dependent on their codimension. In Sec. 3
we consider motion of a point particle in the gravitational field of
the domain wall. The next Sec. 4 is devoted to general description
of the deformation of the domain wall under gravitational collision
and the derivation of the branon wave equation with a source. In
Sec. 5 we construct the retarded solution for the branon wave
equation with the source generated by gravity of the perforating
particle in even and odd dimensions $D>4$. The Sec. 6 is devoted to
the limiting case of the  light-light perforation. Then in Sec. 7 we
consider the case of the domain wall in the four-dimensional bulk
which requires special treatment, and in the last Sec. 8 we discuss
some tentative applications. In the Appendix the evaluation of
typical integrals involved in the calculations  is presented.

\section{Gravity of infinitely thin planar walls in $D$ dimensions}

\subsection{Exact solutions}
For more generality we start with considering an  arbitrary
$p$-brane propagating in $D-$dimensional curved space-time (the
bulk).  We denote the bulk metric as $g_{MN},\;M,N=0,1,2,...
,D-1$, and define the brane world-volume $\cV_{p+1}$  by the
embedding equations
 $   x^{M}=X^{M}(\sigma_\mu),\; $
parameterized by arbitrary coordinates  $\sigma_\mu,  \;
    (\mu=0,...,p)$ on $\cV_{p+1}$ .
The corresponding action in the Polyakov form is a functional of
$X^{M}(\sigma_\mu)$ and the metric $\gamma_{\mu\nu}$ on $\cV_{p+1}$:
\begin{align}\label{Bac}
S_p = -\fr{\mu}{2}\int\left[\vp
  X_\mu^M X_\nu^N g_{MN}\gamma^{\mu\nu}-(p-1)\right]\sqrt{|\gamma|}\,d^{p+1}
  \sigma   \,.
 \end{align}
Here $\mu$ is the brane tension, $X_\mu^M=\pa X^M/\pa\hsp\sigma^\mu$
are the tangent vectors and $\gamma^{\mu\nu}$ is the inverse metric
on $\cV_{p+1}$, $\gamma={\rm det} \gamma_{\mu\nu}$. Also,
$\vak^2\equiv 16\pi G_D$,  the metric signature is  $+--\ldots$ and
our convention for the Riemann tensor is $R^B{}_{NRS}\equiv
\Gamma^B_{NS , R}-\ldots\;$. Variation of (\ref{Bac}) with respect
to $X^M$ gives the brane equation of motion
 \begin{align} \label{em}
\partial_\mu\left(  X_{\nu}^N
g_{MN}\gamma^{\mu\nu}\sqrt{|\gamma|}\right)=\frac{1}{2}\,
g_{NP,M}X^N_{\mu} X^P_{\nu}\gamma^{\mu\nu}\sqrt{|\gamma|}\,,
 \end{align}
which  is  covariant with respect to both the space-time and the
world-volume diffeomorphisms. Variation over $\gamma^{\mu\nu}$
gives the constraint equation
 \begin{align} \label{con} \lbr X_\mu^M X_\nu^N - \fr12\,
\gamma_{\mu\nu}\gamma^{\la\tau} X_\la^M X_\tau^N \rbr g_{MN}
+\fr{p-1}{2}\,\gamma_{\mu\nu}=0\,,
 \end{align}
 whose
solution defines $\gamma_{\mu\nu}$ as the induced metric on
$\cV_{p+1}$:
 \begin{align} \label{ceq} \gamma_{\mu\nu}=X_\mu^M  X_\nu^N g_{MN}{\big
 |}_{x=X}\,.
 \end{align}
 Adding to (\ref{Bac}) the Einstein action with the cosmological
 constant
\begin{align}\label{Eac}
S_E =   -  \frac{1}{\vak^2}\!\int\! \left(R_D+2\Lambda\right)
\,\sqrt{|g|}\; d^D x \,,
 \end{align}
 and varying $S_p+S_E$ with respect to the space-time
metric $g_{MN}$ we obtain Einstein equations
\begin{align}\label{Eeq}
 R_{MN}-\frac12\, g_{MN} R=\frac{\vak^2}{2}\,T_{MN}+\Lambda \hsp g_{MN}
 \end{align}
 with the source term
\begin{align}\label{EMT}
T^{MN}=\mu\int X^M_\mu X^N_\nu
\gamma^{\mu\nu}\;\frac{\delta^D\!\left(x-X (\si)\smph
\right)}{\sqrt{|g|}}\;\sqrt{|\gamma|}\;d^{D-1}\si\,.
 \end{align}

We will be interested in static solutions of the system (\ref{em},
\ref{con}, \ref{Eeq}) for planar branes   described by the linear
embedding functions
 \begin{align}\label{unex}
 X^M =\Si^M_\mu \si^\mu
 \end{align}
 with constant $\Si^M_\mu$. In what follows  we will mostly use the
coordinates $\sigma^\mu$ coinciding with $x^\mu$,  such that
$\Si^M_\mu=\delta^M_\mu$, but in some cases $\sigma^\mu$ will be
also invoked  to avoid confusion.

Consistency of the above coupled system involving singular delta
sources depends on codimension $\tilde d=D-p-1$ of the embedding
of the brane world-volume into the bulk. Strictly speaking, for
$\tilde d\geq 3$ the use of distributions in the  full non-linear
gravity is not legitimate, though the  presence of delta-sources
in classical $p$-brane solutions in supergravities sometimes still
can be detected \cite{Duff:1994an}. The case  $\tilde d=2$ and
$\Lambda=0$, as it is well-known from an example of the cosmic
string in four-dimensional space-time \cite{Vilsh}, is
exceptional: in this case the cylindrically symmetric field
configurations exist for which Einstein equations reduce to
two-dimensional Laplace equation with the delta-source leading to
static locally flat conical transverse space. The case $\tilde d=
1$ (domain wall) is legitimate too, but has a peculiar feature:
for $\Lambda=0$ exact solutions of Einstein equations are
non-static \cite{linear, linear1}. However, static solutions in
this case do exist for some special value of the cosmological
constant $\Lambda<0$, the notorious example being the
Randall-Sundrum metric in $D=5$ \cite{RS, RS1, RS2}. Indeed, with
an ansatz
 \begin{align}\label{drs}
 ds_D^2=\e^{-2F(\bar{z})}\eta_{\mu\nu} dx^\mu dx^\nu-d\bar{z}^2\,,
 \end{align}
the Einstein equations reduce to:
\begin{align}\label{eeqF}
(D-2)F''=\fr{\mu \vak^2}{2}\,\delta(\bar{z})\, , \qquad\quad
(D-1)(D-2)(F')^2+2\Lambda=0 \,,
\end{align}
where a prime denotes the differentiation over $\bar{z}$. This
system has an exact solution provided
\begin{align}\label{lamka}
 \Lambda=-
 \frac{\mu^2 \vak^4 (D-1) }{32\left(D-2\right)}  \,.
\end{align}
Imposing the additional $\mathbb{Z}_2$-symmetry
$F(-\bar{z})=F(\bar{z})$,
 one obtains
\begin{align}\label{rs2coupl}
F=k|\bar{z}|\,,
\end{align}
where
\begin{align}\label{ka}
k=
 \frac{\mu \varkappa^2_D }{4\left(D-2\right)}  \,,
\end{align}
so  an exact solution of Einstein equations with the cosmological
constant (\ref{Eeq}) reads:
\begin{align}\label{mRS}
ds^2=\e^{-2k|\bar{z}|}\eta_{\mu\nu}dx^\mu dx^\nu-d\bar{z}^2.
\end{align}

For comparison, we also give here the time-dependent solution
found by Vilenkin \cite{linear,Vilsh} and Ipser and Sikivie
\cite{IS, IS1} in four dimensions which exists in absence of the
cosmological constant:
\begin{align} \label{vis1}
ds^2=\left(\vp 1-k |z|\right)^2\,dt^2\,-\,\e^{2kt}\left(\vp 1-k
|z|\right)^2(dx^2+dy^2)-dz^2.
 \end{align}
 Here $k$ is a free parameter.

\subsection{Linearized gravity}
Now we pass to  linearized gravity in  Minkowski space-time assuming
$\Lambda=0$ and expanding the metric as
\begin{align}
\label{meka} g_{MN}=\eta_{MN}+ \vak  h_{MN}\,.
\end{align}
All subsequent operations with indices of $h_{MN}$ will be performed
with respect to the Minkowski metric, e.g., $g^{MN}\approx
\eta^{MN}- \vak  h^{MN}$. In the Lorentz gauge
\begin{align}\label{Lorg}
\pa_N h^{MN}=\frac{1}{2}\, \pa^M h, \quad h=h^M_M\;,
\end{align}
the linearized Einstein equations reduce to
\begin{align}
 \Box\, h_{MN}=- \vak
\left( T_{MN}-\frac{1}{D-2}\,  T\, \eta_{MN}  \right)\,,\quad
T=T_M^M\, ,
\end{align}
with $\Box \equiv \partial_M \partial^M$. Consider again an
arbitrary plane unexcited $p$-brane described by the embedding
functions
 (\ref{unex}), choose the coordinates on $\cV_{p+1}$
as $\sigma^0=x^0\equiv t\,,\sigma^i=x^i\,, i=1,\ldots,p$ and denote
the
 coordinate transverse to the brane as $y^k\;, k=1,\ldots,\tilde
 d$. Then the brane stress-tensor $T_{MN}$ will have non-zero only
the components $\mu\,,\nu =0,i
 $ given by
\begin{align}
T_{\mu\nu}=\mu\eta_{\mu\nu}\delta^{\tilde d}({\bf z})\;,
\end{align}
where $\eta_{\mu\nu}$ is Minkowski metric on the brane (and unity in
the case $p=0$), leading to
 \begin{align} \label{brgr1}
ds^2=\left(\vp 1+4k({\tilde d}-2)\Phi_{\tilde
d}\right)\,\eta_{\mu\nu}dx^\mu dx^\nu- \left(\vp1-4k(p+1)
\Phi_{\tilde d}\right)d {z_k}^2 \,.
 \end{align}
Here $\Phi_{\tilde d}$ is the solution of the transverse Poisson
 equation
 \begin{align} \label{lm10}
\Delta_{\tilde{d}}\Phi_{\tilde{d}}(\vec{z})
=\delta^{\tilde{d}}(\vec{z})\,,
\end{align}
which reads explicitly
\begin{align} \label{lm11}
  \Phi_{\tilde{d}}(\vec{z})= \Bigg\{
\begin{array}{cl}
{|z|}/{2} , &\mbox{$\tilde{d}=1$} \\
({2\pi})^{-1}\ln{|\bf z|},&\mbox{$\tilde{d}=2$}\\
- (\tilde{d}-2)^{-1}\Omega_{\tilde{d}-1}^{-1}|\vec{z}|^{(2-\tilde{d}
) }, &\mbox{$\tilde{d}\ge 3$}
\end{array}
 \,,
\end{align}
where $\Omega_{\tilde{d}-1}$ is the volume of the
${\tilde{d}-1}-$dimensional unit sphere in $\tilde{d}-$dimensional
euclidean space:
\mbox{$\Omega_{\tilde{d}-1}=2\pi^{\tilde{d}}/\Gamma(\tilde{d})$}.

\subsection{Domain walls}
In the case $\tilde{d}=1$ the metric (\ref{brgr1}) reads
 \begin{align} \label{dwgr1}
ds^2=\left(\vp 1-2k |z|\right)\,\eta_{\mu\nu}dx^\mu dx^\nu-
\left(\vp1-2k(D-1) |z|\right)d {z}^2 \,.
 \end{align}
This solution generalizes  to arbitrary $D$ the Vilenkin solution
\cite{linear, linear1} of the four-dimensional vacuum linearized
gravity, and, similarly, it can not be regarded as linearization
of any static metric satisfying $D$-dimensional vacuum Einstein
equations. However it can be viewed as linearization of the
Randall-Sundrum type exact solution (\ref{mRS}) of the Einstein
equations with the
 negative cosmological constant (\ref{lamka}). Indeed, the
linearization of (\ref{mRS}) for small $k|\bar{z}|$ reads
\begin{align}\label{mRS1}
 ds^2=\left(\vp 1-2k|\bar{z}|\right)\,\eta_{\mu\nu}dx^\mu
dx^\nu-d\bar{z}^2\,.
\end{align}
It looks different from our solution (\ref{dwgr1}), but in fact
(\ref{mRS1})  does not satisfy the Lorentz gauge condition
(\ref{Lorg}) contrary to (\ref{dwgr1}). Therefore, within the
validity of the linear approximation, the solutions (\ref{mRS1}) and
(\ref{brgr1}) must be related by some coordinate transformation. It
is easy to check that
\begin{align}
\bar{z}=z-\frac{D-1}{2}\,kz^2 \sgn(z)\,,
\end{align}
does this job in the linear order in  $k|z|$. Note that the right
hand side of this equation is continuous at $z=0$, so in the
vicinity of this point  $|\bar z|=|z|$. Therefore, the linearized
domain wall can be viewed as the small distances limit $k|z|\ll 1$
of the exact solution (\ref{mRS}). Note in passing that the required
cosmological constant $\Lambda$ in (\ref{Eeq}) is quadratic in $k$,
so it can be neglected in the linear order in $k$ .
\subsection{Interaction between plane parallel branes}
It is instructive to explore linearized gravitational interaction
between two plane parallel branes $p$ and $\bar p\leq p$, sitting at
some finite distance. We split the space-time coordinates as $
x^M=(t, \vec{x}, \vec{y}, \vec{z}), $ where $ \vec{x}\in
\mathbb{R}^{\bar{p}}, $  $ \vec{y}\in \mathbb{R}^{p-\bar{p}},$ $
\vec{z}\in \mathbb{R}^{\tilde{d}}.$ Let the  first $p$-brane occupy
 the sector $ x^A= (t,\vec{x}, \vec{y}) $ and located at $\bf z=0$ in
the overall transverse space, while the second extends in the
sector $ x^{a}=(t,\vec{x}) $ at the position $\bf
z={\bf{\bar{z}}}$. To extract the effective interaction potential
we start with the action
\begin{align} \label{lm1}
 S_{\rm{int}}=-\frac{\vak}{2}\int \vp h_{MN} \bar{T}^{MN} d^Dx\,,
 \end{align}
where $h_{MN}$ is the linearized metric of the $p$-brane and
$\bar{T}^{MN}$ is the stress-tensor of the $\bar p$ brane (or
vice-versa) and insert as $h_{MN}$  the solution of the
corresponding d'Alembert equation. Using the  scalar Green's
function of the d'Alembert equation
\begin{align} \label{lm6}
\Box_D G(x,x')=\delta^D(x-x')\, ,
\end{align}
we obtain  the bilinear form of the stress-energy tensors
\begin{align} \label{lm4}
S_{\rm int}=-\frac{\vak^2}{2} \int  \vp G(x,x') \!\left(T^{MN}(x) \,
\bar{T}_{MN}(x')- \frac{1}{D-2} \, T(x)\, \bar{T}(x')\right)\dDx
\hsp \dDx'\,.
\end{align}
 Substituting here the corresponding quantities for both branes
  at rest, we find that the integral
(\ref{lm4}) reduces to that over time and the spatial coordinates
$\bf x$ of the $\bar p$-brane, allowing for introduction of the
effective potential $U_{\rm eff}$ per unit volume of the smaller
brane:
\begin{align} \label{lm8} S_{\rm int}=-\int U_{\rm eff}(\vec{\bar{z}})\,dt \smhsp d\vec{x},
\end{align}
which explicitly reads
\begin{align} \label{lm9}  U_{\rm eff}=\frac{\varkappa_D^2\mu\bar{\mu}
(\bar{p} +1
)(\tilde{d}-2)}{2(D-2)}\;\Phi_{\tilde{d}}(\vec{\bar{z}})\,.
\end{align}
Inserting here the transverse potential (\ref{lm11}) we finally
obtain
\begin{align}
U_{\rm eff}=-\frac{\vak^2\mu\bar{\mu}(\bar{p} +1 )}{2(D-2)}\,
\left\{
\begin{array}{cl}
  \bar{z}/2 , &\mbox{$\tilde{d}=1$} \\
0,&\mbox{$\tilde{d}=2$}\\
 {\Omega^{-1}_{\tilde{d}}\,|\vec{\bar{z}}|^{(2-\tilde{d})}}, &\mbox{$\tilde{d}\ge 3$}
\end{array}\right.\quad .
\end{align}

Thus, the character of interaction  depends on codimension of the
embedding of the bigger $p$-brane into the bulk: the potential is
repulsive for $\tilde{d}=1$, there is no force for $\tilde{d}=2$
and it is attractive for $\tilde{d}>2$. This simple picture,
however, holds only in the static case. As we will see, situation
becomes more sophisticated when branes are in motion.

\section{Interaction of domain wall with moving point particle}
Now we wish to consider the system of the gravitationally
interacting domain wall and a moving point particle ($p=0$ brane).
This can be done adding to the sum $S_p+S_E$ the particle action
\begin{align}\label{Bac_part}
S_0 =   - \frac{1}{2} \int \! \left(e\; g_{MN}\dot{z}^M
\dot{z}^N+\frac{m^2}{e}\right) d\tau  \,,
 \end{align}
 where $e(\tau)$ is the ein-bein of the particle world-line and
dots denote derivatives with respect to $\tau$. Varying $S_0$ with
respect to  $z^M(\tau)$ and $e(\tau)$ one obtains the geodesic
equation in arbitrary parametrization
\begin{align}\label{eomp}
\fr{d}{d\tau}\lbr e \smhsp \dot{z}^N g_{MN} \rbr=\fr{e}2 \;
g_{NP,M}\hsp\dot{z}^N \dot{z}^P\, ,
\end{align}
and the constraint equation
\begin{align}
\label{consp} e^2  g_{MN} \dot{z}^M \dot{z}^N=m^2\,.
\end{align}
The corresponding energy-momentum tensor reads
\begin{align}\label{EMT_part}
\bar T^{MN} = \int \fr{e \zt^M \zt^N \delta^D\!\left(x-z(\tau)\smph
\right)}{\sqrt{|g|}}\,d\tau \,.
 \end{align}

 Both the domain wall and the point particle will be treated on an
equal footing in the framework of the linearized gravity on
Minkowski  background. So we expand the total metric similarly to
(\ref{meka}) adding to the metric perturbation $h_{MN}$, which
will be still associated with the brane, the metric perturbation
$\bar h_{MN}$ due to the particle (preserving the notation of sec.
2D):
\begin{align}
\label{meka1} g_{MN}=\eta_{MN}+ \vak  \left(h_{MN}+\bar
h_{MN}\right)\,.
\end{align}
The Lorentz gauge condition (\ref{Lorg}) will be assumed for both
components independently.

To treat the interaction problem in terms of formal expansions in
the gravitational coupling we have now to expand the embedding
functions $X^M(\sigma)$ and $z^M(\tau)$ as well as the Lagrange
multipliers $\gamma_{\mu\nu}$ and $e(\tau)$ in powers of $\vak$ to
the first order, which amounts  to replacing these quantities by
$X^M+\delta X^M $, $z^M +\delta z^M$,$\gamma_{\mu\nu} +\delta
\gamma_{\mu\nu}$ and $e +\delta e$. Here $X^M, z^M, \gamma_{\mu\nu},
e $ will now correspond to free motion in Minkowski space-time,
while $\delta X^M\,,\delta \gamma_{\mu\nu} $ are the perturbations
of the brane variables due to the gravitational field of the
particle $\bar h_{MN}$, and $\delta z^M, \delta e$ are the
perturbations of the particle variables due to the gravitational
field of the brane $ h_{MN}$ (and we omit singular self-interaction
terms).

\subsection{Perturbation of the particle world-line}
The unperturbed domain wall is described by the embedding functions
(\ref{unex}) and the corresponding induced metric is
\begin{align}
 \gamma_{\mu\nu}=\eta_{\mu\nu} \,.
\end{align}
Its gravitational field can be read off from Eq. (\ref{dwgr1}), or
explicitly
 \begin{align} \label{brgr}
h_{MN}=\frac{\vak \mu}2\left(\Xi_{MN}
-\frac{D-1}{D-2}\,\eta_{MN}\right)|z|=\frac{\vak \mu
|z|}{2(D-2)}\;{\rm diag}\hsp(-1,1,...,1,{D-1})\,,
 \end{align}
where $ \Xi_{MN} \equiv \Si_M^\mu\Si_{N}^{\nu} \eta_{\mu\nu}$.

Assuming the particle to move orthogonally to the wall, we
parameterize the unperturbed world-line as
   \begin{align} \label{zzero}
z^M(\tau)= u^M \tau\, ,\quad u^M=\gamma\,(1, 0, ..., 0, v)\,,\quad
\gamma=1/\sqrt{1-v^2}\,.
\end{align}
This  trajectory intersects the domain wall at the moment of proper
time $\tau=0$, the corresponding coordinate time also being zero,
$t=0$. Using (\ref{brgr}) and (\ref{zzero}) in the Eqs.
(\ref{consp}) and (\ref{eomp}) one obtains for $ \delta e$ and $
\delta z^M$ the system of equations
\begin{align}
\label{e1eq}  \delta e=-\frac{ m}{2} \left( \vak\hsp h_{MN}u^M
u^N+2\,\eta_{MN}\hsp u^M \hsp\delta
 \dot{z}^N \vp \right)
\end{align}
and
\begin{align}\label{cucu0}
\fr{d}{d\tau} \left(\vp \delta e \hsp u_M +m\, \delta \dot{z}_M
\right) =-\vak m
 \left(h_{PM,Q}-\frac{1}{2} \,h_{PQ,M}  \right)  u^P u^Q\,,
\end{align}
which upon the elimination of $ \delta e$ gives for $\delta z^M$:
\begin{align}\label{z1eq}
 \bar\Pi^{MN}   \delta\ddot{z}_N=-\vak\hsp
\bar\Pi^{MN}\!\left(h_{PN,Q}-\fr12\,h_{PQ,N}  \right) u^P u^Q \,,
\end{align}
where
\begin{align}\label{pimen}
\bar\Pi^{MN}=\eta^{MN}-u^M u^N
\end{align}
is a projector onto the subspace orthogonal to $u^M$. Let us now
choose the  overall gauge condition
\begin{align}
g_{MN} \dot z^M \dot z^N=1\,,
\end{align} with $z^M$ including the
perturbation. In view of the zero order parametrization assumed
(\ref{zzero}), this amounts to the condition $\delta e=0$, i.e.,
\begin{align}
\label{pgage}  \frac{ m}{2} \left( \vak\hsp h_{MN}u^M
u^N+2\,\eta_{MN}\hsp u^M \hsp\delta \dot{z}^N \vp \right)=0\,.
\end{align}
Going back to eq. (\ref{cucu0}) one has thereby
\begin{align}\label{cucu0_a}
     \delta\ddot{z}_M =-\vak
 \left(h_{PM,Q}-\frac{1}{2}\, h_{PQ,M}  \right)u^P u^Q\,,
\end{align}
 or, in components,
 \begin{align} \label{acce1}
 \delta \ddot{z}^{\,0} =2kv\,
\gamma^2 \;{\rm sgn }(\tau)\,,\quad  \delta\ddot z \equiv
\ddot{z}^{D-1}=k\, (D\gamma^2 v^2+1)\;{\rm sgn } (\tau) \,,
 \end{align}
so, the force is repulsive as expected.

Integrating  (\ref{acce1}) twice with initial conditions $ \delta
z^M(0)=0, \;\delta\zt^M(0)=0$, one has
 \begin{align} \label{acce2}
 \delta z^0   =k v \tau^2\, \gamma^2
\;{\rm sgn }(\tau)\,,\qquad { \delta z }=\frac{1}{2}\, k\tau^2
\left(D\gamma^2 v^2+1\right)\,{\rm sgn }( \tau)\,.
 \end{align}
Substituting  (\ref{acce2}) into (\ref{e1eq}) one can check that the
gauge condition (\ref{pgage})  holds.

Since we are using the perturbation theory, corrections to the
uniform particle motion must be small (assuming appropriate initial
conditions), so the particle always hits the brane and perforates
  it, reappearing on the other side. The
reflection is of course physically possible, if the velocity is not
high enough, but this is beyond the validity of our approximation.
Our aim is to consider in detail what happens when the particle
pierces the brane, so the case of reflection is outside the scope of
the present treatment.

 \subsection{Acceleration discontinuity at the moment of perforation}
According to (\ref{acce2}), the perturbation of the particle
 energy $\delta p^0=m\delta\dot z^0$ and the momentum $\delta
p^z=m\delta\dot z$ have no discontinuity at the location of the
brane $z=0$, but their derivative have. The point-like particle
therefore  perforates the domain wall without loss of the
energy-momentum, but its acceleration is finite and
instantaneously changes sign.   The sign rule in (\ref{acce1})
corresponds to gravitational repulsion, as could be expected in
the case of co-dimension one. Therefore the discontinuity of
acceleration has a simple physical meaning: the repulsive force
changes its sign at the moment of perforation.

For consistency of our perturbative approach we have to ensure that
the first order correction to the particle momentum remains small in
the vicinity of the wall $k|z|\ll 1$ where the linearized
approximation for wall's gravity is valid. Since in the zero order
$z=v\gamma\tau$, from (\ref{acce2}) we find that this is true
indeed:
 \begin{align} \label{deltav}
|{  \delta\zt^0} | \sim k |z|\gamma \ll \gamma =u^0\,.
 \end{align}

\section{Deformation of  domain wall}\label{Branon_wave_equation}
Now we explore perturbations of the domain wall due to gravitational
interaction with the  perforating particle. For this we need to know
the metric perturbation due to the particle. In accordance with the
iterative approach adopted here, we must neglect  the  particle
acceleration in the wall's gravity when we calculate its proper
gravitational field, considering the unperturbed particle
trajectory.
\subsection{Particle gravity}
 In what follows, the stress-tensor and the gravitational field of
the particle will be denoted by bar. The metric perturbation  moving
along the straight line in the Minkowski space satisfies the
equation
\begin{align}\label{barheq}
 \Box_D\, \bar h_{MN}=- \vak
\left( \bar T_{MN}-\frac{1}{D-2}\,  \bar T\, \eta_{MN} \right)\,,
\end{align}
with the source term
\begin{align}
\label{T0mn}  \bar T^{MN}(x)=  m\int u^M u^N
\delta^D\left(x-u\tau\right)\, d\tau\,,
\end{align}
which   has only $t,z-$ components non-zero. Passing to the
$D$-dimensional Fourier-transforms
\begin{align}\label{fur}
 & \bar h_{MN}(x)=\frac1{(2\pi)^D}\int \e^{-iqx}\bar h_{MN}(q)
\,d^Dq\,,\non\\ &\bar T^{MN}(x)=\frac1{(2\pi)^D}\int \e^{-iqx} \bar
T^{MN}(q)\,d^Dq\,,
\end{align}
we obtain from (\ref{barheq}) the retarded solution  in the
momentum representation
 \begin{align}\label{ge_mom}
   \bar{h}_{MN}(q) =\frac{2\pi \vak m \, \delta(qu)}{q^2+i \varepsilon q^0} \left(u_M
u_N-\fr{1}{D-2}\,\eta_{MN}\right).
  \end{align}
In the coordinate representation we find (for $D\geq 4$):
\begin{align} \label{hpart}
 \bar{h}_{MN}(x)=-\fr{\vak
\,m\Gamma\left(\fr{D-3}{2}\right)}{4\pi^{\fr{D-1}{2}}}
 \left(u_M
u_N-\fr{1}{D-2}\,\eta_{MN}\right)\fr{1}{[\gamma^2(z-v
t)^2+r^2]^{\fr{D-3}{2}}}\,,
 \end{align}
where $r= \sqrt{\delta_{ij} x^i x^j}$ is the radial distance on the
wall from the perforation point. This is just the Lorentz-contracted
$D$-dimensional Newton  field of the  uniformly  moving particle.
\subsection{Perturbation of the wall}
Perturbations of the Nambu-Goto branes in external gravitational
field were expensively studied in the literature, see e.g.
\cite{Gu, Gu2}. On the Minkowski background the derivation is
particularly simple. First, from Eq. (\ref{ceq}) we find the
perturbation of the induced metric
 \begin{align}
  \delta \gamma_{\mu\nu}=2\hsp \delta^M_{(\mu}\hsp \delta\nhsp \smhsp X^N_{\nu)}\eta_{MN}+
 \vak \bar h_{MN}\Sigma^{M}_{\mu}\Sigma^{N}_{\nu} \,,
 \end{align}
where brackets denote symmetrization over indices with the factor
$1/2$. Then linearizing the rest of the Eq. (\ref{em}), after some
rearrangements one obtains the following equation for deformation of
the wall:
 \begin{align}  \label{pisk}
 {\Pi}_{MN}\;\Box_{D-1}\;\delta\nhsp \smhsp
X^N= {\Pi}_{MN}\;J^N\,, \quad  {\Pi}^{MN} \equiv \eta^{MN}-
\Sigma^{M}_{\mu}\Sigma^{N}_{\nu} \eta^{\mu\nu}\,,
 \end{align}
 where $\Box_{D-1} \equiv \partial_{\mu} \partial^{\mu}$ and $ {\Pi}^{MN} $ is the
projector onto the (one-dimensional) subspace orthogonal to
$\cV_{D-1}$. The source term   in  (\ref{pisk}) reads:
 \begin{align} \label{JN}
 J^N=   \vak \, \Sigma_P^\mu \,\Sigma_Q^\nu\,
\eta_{\mu\nu} \left(\frac{1}{2} \, \bar h^{PQ,N} -\bar
h^{NP,Q}\right)\Bigg|_{\!z=0}\! .
 \end{align}
Using the aligned coordinates on the brane
$\sigma^{\mu}=(t,\mathbf{r}), $ we will have
$\delta^{M}_{\mu}=\Sigma^{M}_{\mu}$, so the projector $
{\Pi}^{MN}$ reduces the system (\ref{pisk}) to a single equation
for $M=z$ component. Thus only the $z$-components  of $\delta\nhsp
\smhsp X^M$ and $J^M$ are physical. Generically, the transverse
coordinates of the branes can be viewed as Nambu-Goldstone bosons
(branons) which appear as a result of spontaneous breaking of the
translational symmetry \cite{KuYo}. These are coupled to gravity
and matter on the brane in the brane-world models  via the induced
metric (for a recent discussion see \cite{Bu, Bu2}). In our case
of co-dimension one there is only one such branon. The remaining
components of the perturbation $\delta\nhsp \smhsp X^M$ can be
removed by suitable transformation of the coordinates on the
world-volume, so $\delta\nhsp \smhsp X^\mu=0$ is nothing but the
choice of gauge. Note that in this gauge the perturbation of the
induced metric $\delta \gamma_{\mu\nu}$ does not vanish, as it was
for the perturbation of the particle ein-bein $e$.

 Denoting the physical component as $\Phi(\sigma^\mu) \equiv
\delta\nhsp \smhsp X^z$ we obtain the branon $(D-1)$-dimensional
wave equation:
 \begin{align} \label{NGEQ}
\Box_{D-1} \Phi(\si^{\mu})=J(\si^{\mu}),
  \end{align}
  with  the source term $J \equiv J^z$.
Substituting (\ref{hpart}) into the eq.\,(\ref{JN}) we obtain the
source term for the branon:
 \begin{align} \label{jxb}
 J( \sigma )=-\vak \left[\frac{1}{2}\,  \eta_{{\mu\nu}}\bar{h}^{\hsp \mu\nu,z}-\bar{h}^{\hsp  z\hsp  0,0}\right]_{z=0} =-  \fr{\la vt}{[\gamma^2 v^2
t^2+r^2]^\fr{D-1}{2}}\,,\end{align} where\begin{align}
\la=\fr{\vak^2 m\gamma^2\Gamma\left(\fr{D-1}{2}
\right)}{4\pi^{\fr{D-1}{2}}}\left( \gamma^2v^2 +\fr{1}{D-2}\right) .
 \end{align}

\subsection{Nature of  singularity of  the source}

The source-current has some peculiar features. It is a smooth
function of $r,\,t$ except for the point $r=0$ where it has
singularity at the moment $t=0$ of perforation. This singularity is
due to singular nature of the Coulomb field of the point-like
particle. Of course  the linearized gravity theory can not be
trusted as description of the gravitational field near the
singularity. Nevertheless, we know that the Fierz-Pauli theory {\em
can} be used   to consistently describe gravitational interaction of
point masses treated in terms of distributions. Since the above
singularity is essentially due to the point-like nature of the
particle stress-tensor, one can hope to be able to describe the
whole situation  in terms of distributions too.

Consider  the branon wave equation with the source treated as
distribution. It is easy to see, that in the limit $t\to 0$ or $v\to
0$ ($\gamma \to 1$), the source  exhibits properties of the $n=(D-2)
\,$-dimensional delta-function. Denoting $\al=\gamma v t$, one has:
 \begin{align}
\lim_{\al\to \pm 0}
\frac{\al}{(r^2+\al^2)^\fr{n+1}{2}}=\Bigl\{\begin{array}{cc} 0,\;&
 \quad r\neq 0\,, \\ \pm\infty, & \quad r=0\,.
\end{array}
 \end{align}
The integral of $J(x)$ over the $n$-dimensional space is
$\al$-independent (up to the sign) and finite:
 \begin{align}
\int J(x)\, d^{\hsp D-2}\nhsp x=-\la \Omega_{n-1}
\int\limits_0^{\infty}
\frac{\al\,r^{n-1}}{(r^2+\al^2)^\fr{n+1}{2}}\, dr=-\frac{
\pi^\fr{n+1}{2}\la }{\Gamma\left( \frac{n+1}{2}\right) \gamma}\,
\sgn (\al)\,,
 \end{align}
so the integrand is proportional to the $n\,$-dimensional
delta-function. Since $\sgn(\al)=\sgn(t)$, we get therefore:
 \begin{align}  \lim_{vt\to \pm 0}
J(x)=-\frac{ \pi^\fr{ n+1 }{2}\la
 }{\Gamma\left(
\frac{n+1}{2}\right)\gamma}\:\sgn(t)\,\delta^n(\r)\,.
 \end{align}
It is worth noting that this limiting distribution will be the same
either we consider {\em time} in the close vicinity of the
perforation moment $t\to 0$ for any velocity $v$ of the mass $m$, or
if we consider the limit of the {\em small velocity} $v\to 0$. In
the latter case (the quasi-static perforation) this limit holds for
sufficiently large $t$, and since the coefficient $\la$ remains
finite as $v\to 0$, the point-like source at the right hand side of
the branon field equation (\ref{NGEQ}) may be attributed to an
effective branon ``charge'', or the perforation charge.

This notion allows us to better understand the difference between
two cases. The first is  the static point mass sitting on the brane
eternally. Then, coming back to the Eq. (\ref{pisk}) for brane
perturbations, we find that the source term at the right hand side
will be zero (i.e. there is no perforation charge in absence of
perforation).  On the contrary, if  the  perforation takes place
even adiabatically slowly, the branon charge is non-zero. Indeed, in
the limit $v\to 0$ we will have a point-like source of the branon
field:
 \begin{align}  \lim_{v\to 0} J(x)= Q_{B}\,\delta^n(\r)\,,
 \end{align}
 where an effective branon charge is given by
\begin{align}  \label{QNG} Q_{B}=-k\, \sgn(t)\,.
 \end{align}
This ``charge'' is a manifestly non-conserved quantity, changing
sign at the moment of perforation. For an observer on the brane
the perforation therefore looks like a sudden shake, and, as we
will see in the next section, the corresponding branon field will
be not a static Coulomb field, but an expanding wave.

\section{Constructing the retarded solution}\label{Ret_sol_br}

In view of causality, it is reasonable to construct the retarded
solution of the branon wave equation (\ref{NGEQ}) generated by the
source. This can be done using the standard retarded Green's
function on a $(D-1)$-dimensional flat manifold:
 \begin{align}
G_{\rm ret}(x-x')=-\frac{1}{(2\pi)^{D-1}}\int \frac
{\e^{-ik(x-x')}}{k_{\mu}k^{\mu}+2i\ep k^0}\, d^{D-1} k\, ,
 \end{align}
 satisfying
\begin{align}
\Box_{D-1} G_{\rm ret}(x-x')= \delta^{D-1}(x-x')\,.
 \end{align}
Parameterizing the $(D-1)$-dimensional wave-vector as
$k^\mu=(\om,\k)$ and denoting $k=|\mathbf{k}|$  (not  to be
confused with  $k$ in the domain wall metric) we present the
retarded solution of (\ref{NGEQ}) as
 \begin{align} \label{Fig}
\Phi(x^{\mu})=-\frac1{(2\pi)^{D-1}}\int \frac {\e^{-ik x
}}{\omega^{2}-k^2+2i\ep \om}\,J(k^{\mu}) \:d^{D-1} k\,,
 \end{align}
where $J(k^{\mu})$ is the Fourier-transform of the
source\footnote{Computation is presented in the Appendix
\ref{app2}, the Eq.(\ref{Jk}).} (\ref{jxb}):
 \begin{align}\label{Jka}
J(k^{\mu})=-\frac{2\pi^{\fr{D-1}{2}}\la\smhsp }{\gamma
\Gamma\left(
\fr{D-1}{2}\right)}\frac{i\om}{\gamma^2v^2k^2+\om^2}\,.
\end{align}

 Integration  in (\ref{Fig}) is
straightforward, but it is worth  giving here  some   details in
order to show  the origin of two physically different components of
the solution. Substituting (\ref{Jka}) into (\ref{Fig}) we pass to
  spherical coordinates in the spatial sector of the momentum
space  $d^{D-1}k=\,k^{D-3}\,dk\, d\om \,d\Omega_{D-3}$. Using
 $\mathbf{kr}=kr \cos\theta$, we integrate  over the angles:
 \begin{align}\label{totangint_a}
\int \e^{\pm i z\cos \theta}d\Omega_{n}=\fr{(2\pi)^{
\frac{n+1}{2}}}{ {z}^\fr{n-1}{2}} J_{\fr{n-1}{2}}(z)\,,
 \end{align}
(derivation is given in (\ref{totangint})), and   split the
integrand into the sum of three terms:
 \begin{align}\label{Phi}
\Phi= \frac{2^{-D/2}\hsp
i\hsp\la}{\sqrt{\pi}\gamma^3\Gamma\left(\fr{D-1}{2} \right)
r^{\fr{D-4}{2}}}\!\int\limits_{0}^{\infty} \! dk
\,J_{\frac{D-4}{2}}(k r)\,{k}^{\frac{D-6}{2}} \!\!
\int\limits_{-\infty}^\infty \! d\om\, \e^{-i\om t} \! \left(
\frac1{\om-k+i\ep}+\frac1{\om+k+i\ep}-\frac{2\om}{\om^2+\gamma^2v^2k^2}\right).
 \end{align}
Note that the first two terms in the bracket correspond to solution
of the homogeneous equation, while the last is related to the
source. All the three integrals over $\om$ can be  evaluated by the
contour integration:
 \begin{align}\label{Phitt}
 &\int\limits_{-\infty}^\infty \frac
 {\e^{-i\om t}}{\om\pm k+i\ep}\, d\om =-2i\pi\theta(t)\,\e^{\pm i
 kt}\,,\non\\
& \int\limits_{-\infty}^\infty \frac
 {\om \,\e^{-i\om t}}{\om^2+\gamma^2v^2 k^2}\,
 d\om =-i\pi\,\sgn(t)\, \e^{- k\gamma v|t|}\,,
 \end{align}
where $\theta(t)$ is the Heaviside  function and $\sgn(t)$ is the
sign function. So we obtain:
 \begin{align}\label{Phi11}
\Phi=  \Phi_{\ah}+\Phi_{\smhsp\bh}\,, \quad \Phi_{\ah}
 \equiv - \Lambda \,\sgn\nhsp(t)I_{\ah}\,,\quad \Phi_{\smhsp\bh}
\equiv 2\, \Lambda  \, \theta(t)I_{\bh}\,,\quad \Lambda \equiv
\frac{ \sqrt{\pi}
 \,\la}{2^{ \frac{D-2}{2}} \gamma^3\Gamma\left(\fr{D-1}{2}
 \right)}\,,
\end{align}
where
\begin{align}\label{Phi13a}
  & I_{\ah}(t,r) = \frac{1}{r^{\fr{D-4}{2}}} \int\limits_{0}^{\infty} \! dk
\,J_{\frac{D-4}{2}}(k r)\,{k}^{\frac{D-6}{2}} \, \e^{- k\gamma
v|t|}\,,
\\  &
 I_{\bh}(t,r)= \frac{1}{r^{\fr{D-4}{2}}} \int\limits_{0}^{\infty} \!
dk(t,r)J_{\frac{D-4}{2}}(k r)\,{k}^{\frac{D-6}{2}} \,\cos \, kt\, .
\label{Phi13b}
 \end{align}
Being averaged over the time of collision, the first term
$\Phi_{\ah}$ (''antisymmetric'') in  (\ref{Phi11})  vanishes;
furthermore, it is
 suppressed by the factor $\gamma$ in the exponent for an ultrarelativistic
collision. On the contrary, the second term $\Phi_{\smhsp\bh}$
(''branon'') starts to be active at the moment of perforation and
remains non-zero afterwards. Note that the quantities   $I_{\ah}$
and $I_{\bh}$ are defined as time-symmetric. Thus we have two type
of integrals over $k$ involving Bessel functions of an argument
$kr$. The first, $ \Phi_{\ah}$, can be integrated for general $D$ in
terms of hypergeometric function:
\begin{align}\label{hypergeom}
\Phi_{\ah} =  \frac{
 \la }{2  \gamma^3\Gamma\left(\fr{D-1}{2} \right)
r^{ D-3}  }\,  \left[  \gamma v t  \hsp \Gamma \!\left(
\!\fr{D-3}{2}\right)   {}_2 {}F_1\!\left(  \fr12\, , \fr{D-3}{2}\,
 ;  \fr{3}{2}\, ;  -\fr{\gamma^2 v^2
t^2}{r^2}\right)\! -\!\fr{\sqrt{\pi} r}{2}\, \sgn(t)\, \Gamma \!
\left(\nhsp \fr{D-4}{2}  \right) \! \right]\,,
\end{align}
while for $\Phi_{\smhsp\bh}$ there is no universal formula.
 Taking into account the recurrence relation for Bessel functions
\begin{align}\label{bessel}
\left( \frac{1}{z} \frac{\partial}{\partial z}
\right)\frac{J_{\nu}(z)}{z^{\nu}}=-\frac{J_{\nu+1}(z)}{z^{\nu+1}}\,
,
\end{align}
  the  integrals of both types can be obtained by consecutive
differentiation over $r$,  taking the lowest dimensions $D=5$ and
$D=6$ as generating.

\subsection{Odd $D$}
Consider first the case $D=5$. For arbitrary $(t,r)$ we start
directly with (\ref{Phi13a}) substituting $J_{1/2}(z)=\sqrt{2/\pi
z}\sin\,z$ :
\begin{align}\label{hyyh}
  I_{\ah} =\int\limits_0^\infty \frac{J_{1/2}(k r)}{\sqrt{kr}}\, \e^{-k\gamma v |t|}d k=\frac{1}{\sqrt{2\pi
  }\, r} \left(\pi -2\, \arctan \frac{\gamma v |t|}{r} \right) =\frac{2}{\sqrt{2\pi
  }\,r } \,\arctan \frac{r}{\gamma v |t|}\, ,
\end{align}
which covers all values of $t$ and $r$. Then, using the recurrence
relation (\ref{bessel}) one obtains for general odd $D\geq 5$:
\begin{align}\label{hyyhwdwdew}
 & I_{\ah}  =  \frac{2}{\sqrt{2\pi
  } } \left(- \frac{1}{r} \frac{\partial}{\partial r}
\right)^{\frac{D-5}{2}} \frac{1}{r} \, \arctan \frac{r}{\gamma v
|t|}\, , \non\\& \Phi_{\ah}=- \frac{2\Lambda}{\sqrt{2\pi
  } } \left(- \frac{1}{r} \frac{\partial}{\partial r}
\right)^{\frac{D-5}{2}} \frac{1}{r} \, \arctan \frac{r}{\gamma v
|t|}\,.
\end{align}
The  second type  integral  (\ref{Phi13b})  in five dimensions is
 \begin{align}\label{kotchergah0}
 I_{\bh}  = \frac{2}{\sqrt{2\pi
  }r} \int\limits_{0}^\infty \fr{\sin kr}{k}\,\cos kt \; dk =
 \frac{\sqrt{2\pi
  }}{4\,r}  \left[\vp\hsp \sgn(r+t)+\sgn(r-t)\right]\,.
\end{align}
Multiplying by the identity $1=\theta(t)+\theta(-t)$, for $t>0$ one
has $\theta(t) \, \sgn(r+t) \equiv 1$ and furthermore
$\sgn(r+t)+\sgn(r-t)=2\hsp \theta(r-t)$, while for $t<0$ one has
$\theta(t) \, \sgn(r-t) \equiv 1$ and $\sgn(r+t)+\sgn(r-t)=2\hsp
\theta(r+t)$. Thus we deduce
\begin{align}\label{kotchergah1}
& I_{\bh}  =  \frac{\sqrt{2\pi
  }}{2\,r}
\,\theta(r-|t|)\,, \non \\& \Phi_{\smhsp\bh}= \frac{\sqrt{2\pi
  }\Lambda }{ r} \,\theta(t)
\,\theta(r-t)\,,
\end{align}
which is the spherical shock wave starting at the moment of
perforation moving outward. Applying again the recurrent relation
(\ref{bessel}) one obtains for general odd $D\geq 5$:
\begin{align}\label{hyyhw2}
 & I_{\bh}  =   \frac{\sqrt{2\pi
  }}{2 } \left(- \frac{1}{r} \frac{\partial}{\partial r}
\right)^{\frac{D-5}{2}} \frac{\theta(r-|t|)}{r}\,,
\non\\&\Phi_{\smhsp\bh}= \sqrt{2\pi
  }\Lambda   \,\theta(t) \left(- \frac{1}{r} \frac{\partial}{\partial r}
\right)^{\frac{D-5}{2}} \,\frac{\theta(r-t)}{r}  \,.
\end{align}

At the moment of perforation $t=0$ the $ \Phi_{\bh}$ part of the
solution has a jump at any point on the wall outside the
perforation point. Its value is obtained  by differentiation of
$1/r$, keeping $\theta(r-t)$ unchanged:
 \begin{align}\label{jump_odd}
 \delta  \Phi_{\smhsp\bh}  =  \sqrt{2\pi}  \,  \Lambda  \left(- \frac{1}{r} \frac{\partial}{\partial r}
\right)^{\frac{D-5}{2}} \frac{1}{r}= \frac{\sqrt{2\pi} \, \Lambda
(D-6)!!}{r^{D-4}}\,.
\end{align}

\subsection{Even $D$ }
Now we start with  $D=6$. The integral (\ref{Phi13a}) gives:
\begin{align}\label{hyyhrffr} I_{\ah}
=\frac{1}{r}\int\limits_0^\infty {J_{1}(k r)} \,\e^{-k\gamma v
|t|}d k=\frac{1}{r^{2}}\left(
\vphantom{\frac{d}{d}}\right.1-\frac{\gamma v |t|}{ \sqrt{\gamma^2
v^2 t^2+r^2}}\left. \vphantom{\frac{d}{d}}\right),
\end{align}
while the second integral (\ref{Phi13b}) is
 \begin{align}\label{jaj1}
I_{\bh}= \frac{1}{r}\int\limits_0^\infty  J_1(k r) \cos(k t)\,d k
=
\frac{\theta(r-|t|)}{r^2}-\frac{\theta(|t|-r)}{\sqrt{t^2-r^2}(|t|+\sqrt{t^2-r^2})}
\, .
\end{align}
Applying the recurrence relation (\ref{bessel}) we obtain for
generic even $D\geq 6$:
\begin{align}\label{hy5}
I_{\ah} = \left(- \frac{1}{r} \frac{\partial}{\partial r}
\right)^{\frac{D-6}{2}} \left[ \frac{1}{r^{2}}\left(
\vphantom{\frac{d}{d}}\right. 1-\frac{\gamma v |t|}{
\sqrt{\gamma^2 v^2 t^2+r^2}}\left.
\vphantom{\frac{d}{d}}\right)\right] \, ,
\end{align}
and
 \begin{align}\label{jaj3}
I_{\bh}= \left(- \frac{1}{r} \frac{\partial}{\partial r}
\right)^{\frac{D-6}{2}}
 \left(\frac{\theta(r-|t|)}{r^2}-\frac{ \theta(|t|-r)}{\sqrt{t^2-r^2}(|t|+\sqrt{t^2-r^2})}\right)
\, .
\end{align}

The value of the jump at $t=0$  is given by
 \begin{align}\label{jump_even}
 \delta  \Phi_{\smhsp\bh}  = 2 \Lambda  \left(- \frac{1}{r} \frac{\partial}{\partial r}
\right)^{\frac{D-6}{2}} \frac{1}{r^2}= \frac{2 \Lambda
(D-6)!!}{r^{D-4}}\, , \quad D=\text{even}.
\end{align}

\subsection{Properties of the solution}
The solution obtained has some unexpected features. One could think
that the particle approaching the wall will continuously deform it
outward  through the gravitational repulsion. After perforation
similar deformation could be expected in the opposite direction.
This kind of deformation is present in our solution indeed as the
antisymmetric in time component $\Phi_{\ah}$. However, this
component alone does not satisfy the branon wave equation
(\ref{NGEQ}) at the moment of perforation in the sense of
distributions. Indeed, at the moment of perforation $t=0$ the
function $\Phi_{\ah}$ has a discontinuity for all $r>0$ equal to
\begin{align}\label{hypergeom_jump}
\delta \Phi_{\ah} =  -\frac{
 \la \sqrt{\pi} \Gamma\left(
\!\fr{D-4}{2}\! \right)}{2 \gamma^3\Gamma\left(\fr{D-1}{2} \right)
r^{ D-4}  }\,,
\end{align}
that immediately follows from the second term in the parenthesis
of (\ref{hypergeom}). Physical origin of this jump is obvious: the
force of interaction remains constant while particle approaches
the wall, but it instantaneously changes sign at the moment  of
perforation. To see the details we first act by the $(D-1)-$
dimensional box on the function $I_{\ah}$. Since (\ref{Phi13a})
does not contain stepwise functions, the second time
derivative\footnote{Hereafter dots over $\Phi(r,t)$ and its
constituents denote the derivative over $t$.} will be continuous:
 \begin{align}\label{cic0}
\ddot{I}_{\ah} =  \frac{\gamma^2 v^2}{r^{\fr{D-4}{2}}}
\int\limits_{0}^{\infty} \! dk \,J_{\frac{D-4}{2}}(k
r)\,{k}^{\frac{D-2}{2}} \, \e^{- k\gamma v|t|}\,.
\end{align}
The $n$ -dimensional laplacian of $I_{\ah}$ is acting only on
 the radial variable, so we can replace it by
$$\Delta_{D-2} = \frac{1}{r^{D-3}} \frac{\partial}{\partial r}
\left( r^{D-3} \frac{\partial}{\partial r} \right).$$ Using
(\ref{bessel}) and the identity
\begin{align}\label{bessel_A}
\left(\frac{1}{z} \frac{\partial}{\partial z} \right) \!\left( \vp
z^{\nu}J_{\nu}(z) \right)=z^{\nu-1} J_{\nu-1}(z)\,,
\end{align}
one obtains
 \begin{align}\label{cic1}
 \Delta_{D-2} {I}_{\ah} =- \frac{1}{r^{\fr{D-4}{2}}}
\int\limits_{0}^{\infty} \! dk \,J_{\frac{D-4}{2}}(k
r)\,{k}^{\frac{D-2}{2}} \, \e^{- k\gamma v|t|}\,.
\end{align}
Combining (\ref{cic0}) and (\ref{cic1}) and using the table
integral from \cite{GR}   we obtain:
 \begin{align}\label{cic2}
  \Box_{D-1} {I}_{\ah} =  \frac{ \gamma^2}{r^{\fr{D-4}{2}}}
\int\limits_{0}^{\infty} \! dk \,J_{\frac{D-4}{2}}(k
r)\,{k}^{\frac{D-2}{2}} \, \e^{- k\gamma
v|t|}=\frac{\Gamma\!\left(\frac{D-1}{2}
\right)}{2^{\frac{2-D}{2}}\sqrt{\pi}}\frac{ \gamma^3
v|t|}{[\gamma^2 v^2 t^2+r^2]^\fr{D-1}{2}}\: .
\end{align}

Now apply the box to the antisymmetric part. Since ${I}_{\ah}$ is
time-symmetric,  then $\dot{I}_{\ah}$ is time-antisymmetric, so
\emph{in the sense of distributions} $\dot{I}_{\ah}\!\nhsp\left.
\vphantom{d^k_K}\right|_{\smhsp t=0}=0$, thus  $\delta(t)
\hsp\dot{I}_{\ah}=0 $. Hence
$$\Box_{D-1}{ \Phi}_{\ah}=-\Lambda \,\sgn\nhsp(t) \hsp\Box_{D-1} {I}_{\ah}
-2\smhsp \Lambda\, \delta'(t) \hsp {I}_{\ah}\!\nhsp\left.
\vphantom{d^k_K}\right|_{\smhsp t=0}\,.$$ The first term comes from
(\ref{cic2}), while the second peaks up the jump of ${\Phi}_{\ah}$:
 \begin{align}\label{cic3}
{I}_{\ah}\!\nhsp\left. \vphantom{d^k_K}\right|_{\smhsp t=0}
=-\frac{\delta \Phi_{\ah}}{2} \, ,
\end{align}
 with $\delta \Phi_{\ah}$ given by
(\ref{hypergeom_jump}). Therefore acting on $\Phi_{\ah}$ by the box
operator we will get the extra term at the right hand side, namely
 \begin{align}\label{jajcc3_mod}
 \Box_{D-1}{ \Phi}_{\ah} =- \frac{\la vt}{[\gamma^2 v^2
t^2+r^2]^\fr{D-1}{2}}+ \delta{ \Phi}_{\ah}\, \delta'(t) \, ,
\end{align}
Since the delta-derivative term is not present in the source of the
branon equation (\ref{NGEQ}), this function is not the full solution
yet.

The second part of the solution $\Phi_{\smhsp\bh}$ just compensates
the delta-derivative term. In fact its presence in the full
deformation was discovered by careful evaluation of the retarded
solution. It is instructive first to act by the box operator on the
function $I_{\bh}$. Differentiating  separately (in the integral
form (\ref{Phi13b})) on time and spatial coordinates one finds,
using the Fourier-transforms (\ref{Phi}), (\ref{Phi11}) and
(\ref{Phi13b}) in any dimension
 \begin{align}\label{jajcc00}
\ddot{I}_{\bh}= \Delta_{D-2} {I}_{\bh} =- \frac{1}{r^{\fr{D-4}{2}}}
\int\limits_{0}^{\infty} \! dk \,J_{\frac{D-4}{2}}(k
r)\,{k}^{\frac{D-2}{2}} \,\cos \, kt\,= -\frac{2^{\frac{D-2}{2}}
\sqrt{\pi}\, t \, |t^2-r^2|^{-\frac{D-1}{2}}}{  \Gamma
\left(-\frac{D-3}{2} \right)} \,,
\end{align}
so it is a solution of the homogeneous wave equation
 \begin{align}\label{jajcc0}
 \Box_{D-1} {I}_{\bh} =0\,.
\end{align}
Note that in odd dimensions the function (\ref{jajcc00}) can be
expressed by derivatives of the delta-function by virtue of the
distributional property \cite{Gelf} $$\lim_{\lambda \to -n}
\frac{|x|^{\lambda}}{\Gamma(\lambda+1 )}=\delta^{(n-1)}(x), \quad
n\in \mathbb{N}.$$ Using this we obtain for the second time
derivative
 \begin{align}\label{jajcc1}
\ddot{I}_{\bh}=  - 2^{\frac{D-2}{2}} \sqrt{\pi}\, t \,
\delta^{\left(\frac{D-3}{2}\right)} (t^2-r^2).
\end{align}

However,  ${I}_{\bh}$ represents the solution part only for $t>0$,
so acting by  box on  $\Phi_{\smhsp\bh}$ we obtain (again using the
$t=0$ limit in (\ref{hypergeom})):
\begin{align}\label{jajcc2}
\Box_{D-1}{\Phi}_{\smhsp\bh}=  2 \Lambda \,
\delta'(t)\hsp{I}_{\bh} \!\nhsp\left.
\vphantom{d^k_K}\right|_{\smhsp t=0}=  \frac{
 \la  \, \sqrt{\pi} \, \Gamma \left(
 \fr{D-4}{2}  \right) }{ 2\gamma^3\Gamma\left(\fr{D-1}{2} \right)
r^{ D-4}  }\, \delta'(t)\,   .
\end{align}
The right hand side exactly compensate the undesirable
delta-derivative term in (\ref{jajcc3_mod}). According to
definitions
 (\ref{Phi13a},\ref{Phi13b}), the  $t=0$ limits $I_{\bh}$ and $I_{\ah}$
coincide:
 \begin{align}\label{Phi13cn}
I_{\bh}\!\nhsp\left. \vphantom{d^k_K}\right|_{\smhsp
t=0}=I_{\ah}\!\nhsp\left. \vphantom{d^k_K}\right|_{\smhsp
t=0}=\frac{1}{r^{\fr{D-4}{2}}} \int\limits_{0}^{\infty} \! dk
\,J_{\frac{D-4}{2}}(k r)\,{k}^{\frac{D-6}{2}} \, ,
 \end{align}
so $I_{\bh}\!\nhsp\left. \vphantom{d^k_K}\right|_{\smhsp t=0}$ is
also given by (\ref{cic3}). So we  conclude that $\delta
\Phi_{\smhsp\bh}=- \delta \Phi_{\ah}$ and the total $\Phi(r,t)$ is
\emph{continuous} at $t=0$ for any $r>0$.

Another split is also useful: both expressions   (\ref{hy5}) and
(\ref{hyyh}) for $I_{\ah}$ (with the corresponding differentiation
for higher odd dimensions) contain two terms -- one is reminiscent
of the static contribution computed in (\ref{jump_even}):
 \begin{align}\label{les_freles_roseaux}
I_{\ah}  \equiv I_{\ah}^{\rm stat} + I_{\ah}^{\rm dyn} \,
,\quad\quad I_{\ah}^{\rm stat}=\frac{2^{\frac{D-6}{2}}
\Gamma\!\left( \frac{D-4}{2}\right)}{r^{D-4}}\,,
 \end{align}
and another (``dynamical'')   vanishing at $t=0$   represents the
hypergeometric part in (\ref{hypergeom}):
\begin{align}\label{hypergeom_dyn}
I_{\ah}^{\rm dyn} = - \frac{2^{ \frac{D-4}{2}}
   \Gamma \left( \!\fr{D-3}{2}\right) }{
 \sqrt{\pi}}\,  \frac{ \gamma v |t| }{r^{ D-3}  } \; {}_2 {}F_1\!\left(  \fr12\,, \fr{D-3}{2}
\,; \fr{3}{2}\,;-\fr{\gamma^2 v^2 t^2}{r^2}\right)  \,.
\end{align}
In this split the component $I_{\ah}^{\rm stat}$ satisfies the
static Poisson equation:
 \begin{align}\label{les_freles_roseaux_2}
\Delta_{D-2} I_{\ah}^{\rm stat}
=-(2\pi)^{\frac{D-2}{2}}\,\delta^{D-2}(\mathbf{r})\,,
 \end{align}
where  $\mathbf{r}$ is a spatial coordinate on the wall. Such a
decomposition will be useful in constructing  solution in four
dimensions.

\subsubsection{Other limits}
Some other limiting cases will be useful:

\vspace{0.5em}  \textbf{$\boldsymbol{r=0,\, t}$\:fixed.} First
consider the values of $\Phi$ at the perforation point $r=0$ for
any time except the perforation moment.  In the Eqs.(\ref{Phi13a},
\ref{Phi13b}) one makes use
 \begin{align}\label{bessel_zero}
 x^{-n}  J_n(x) =  \frac{1}{2^n
 \Gamma(n+1)}\,\left(1+\mathcal{O}(x^2)\right)
 \end{align}
to obtain
 \begin{align}\label{sder}
  & I_{\ah}(t,0) =\frac{1}{2^{\frac{D-4}{2}} \Gamma( \frac{D-2}{2})}   \int\limits_{0}^{\infty} \! dk
\, {k}^{D-5} \, \e^{- k\gamma v|t|}=\frac{
\Gamma(D-4)}{2^{\frac{D-4}{2}} \Gamma( \frac{D-2}{2})}  \,
\frac{1}{(\gamma v |t|)^{D-4}}\,, \\ &
I_{\bh}(t,0)=\frac{1}{2^{\frac{D-4}{2}} \Gamma( \frac{D-2}{2})}
\int\limits_{0}^{\infty} \! dk \, {k}^{D-5} \,\cos \, kt
=\frac{1}{2^{\frac{D-4}{2}} \Gamma( \frac{D-2}{2})}\times
\left\{%
\begin{array}{ll}
    (-1)^{\frac{D-5}{2}}\,\pi\, \delta^{\left(\frac{D-5}{2}\right)}(t), & \hbox{$D=$odd } \\
   \ds  (-1)^{\frac{D-4}{2}}\frac{\Gamma(D-4)}{|t|^{D-4}}, & \hbox{$D=$even} \\
\end{array}%
\right. \, .
 \end{align}

\begin{figure}
\begin{center}
\subfigure[
]{\raisebox{0pt}{\includegraphics[width=8.2cm]{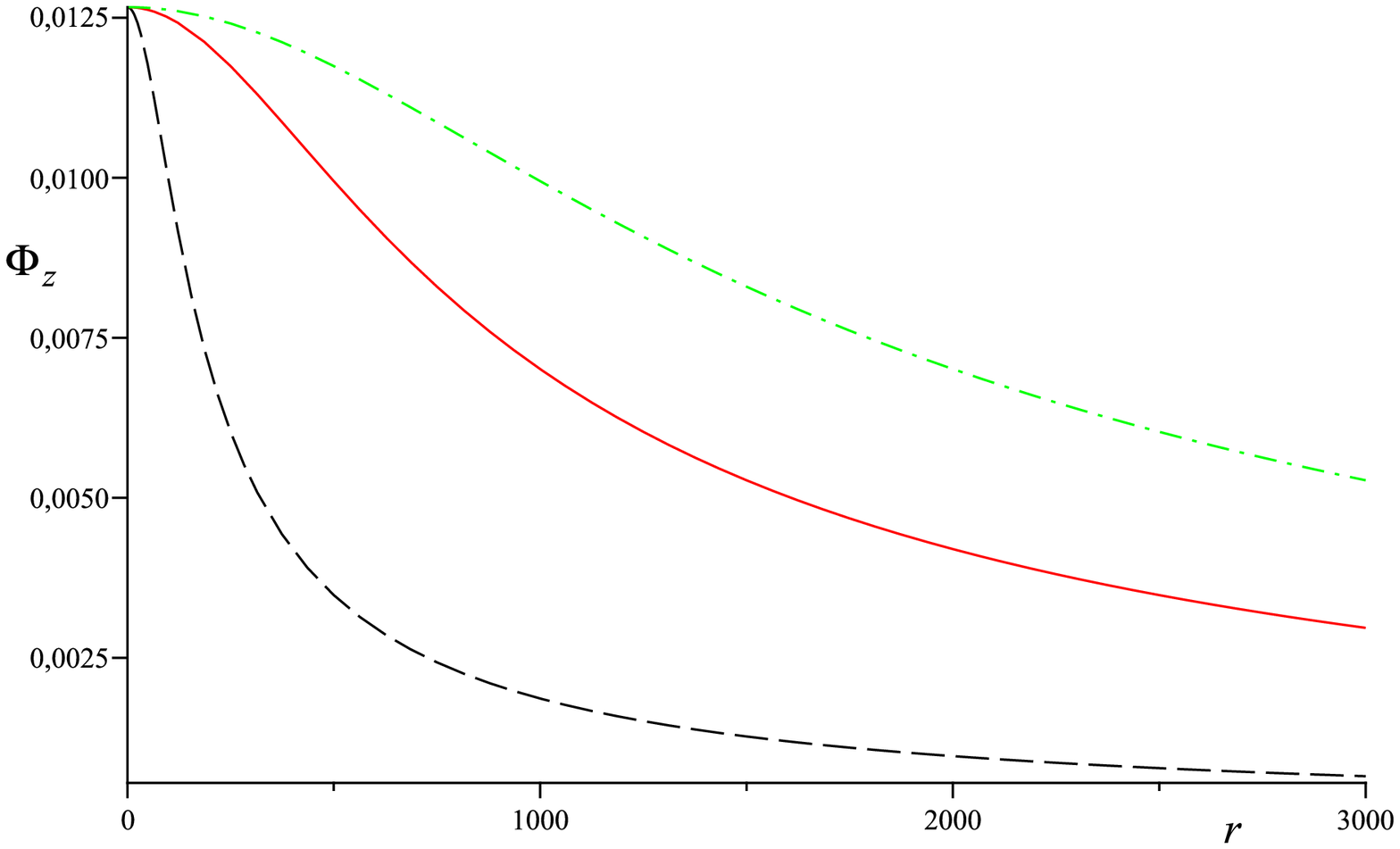}\label{5Dplot_Phiz_r}}}
\subfigure[
]{\includegraphics[width=8.2cm]{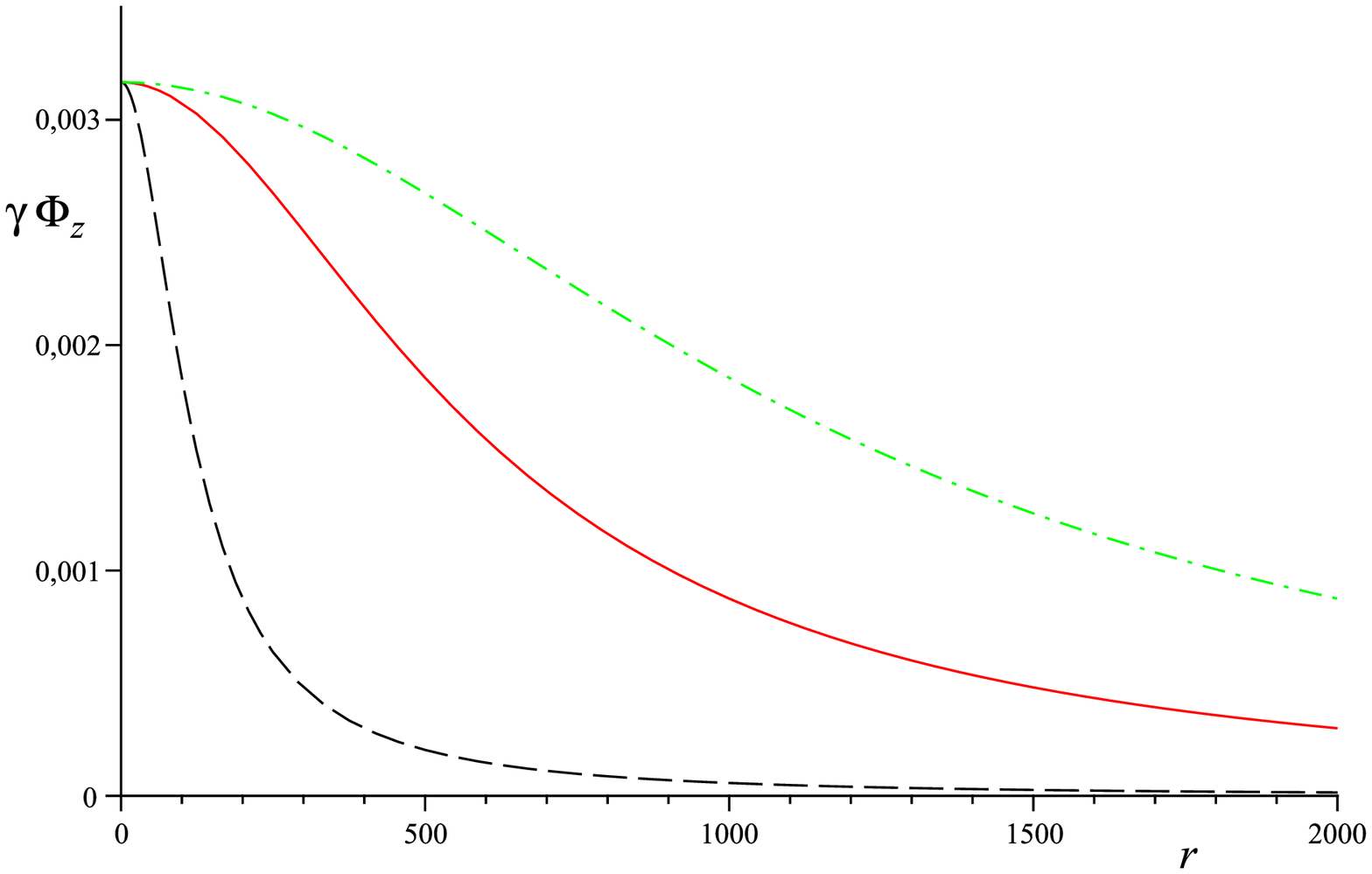}\label{6Dplot_Phiz_r}}
\end{center}
\caption{Behavior of $\Phi_{\ah}$  in five (a) and six (b)
dimensions for fixed value of $t>0$ in $t=1$ scale and units
$\vak=m=1$,   for  $ \gamma=100$ (black, dashdotted), 500 (red,
solid) and 1000 (green, dashed) } \label{56Dplot_Phiz_r}
\end{figure}

As a function of $\gamma$, $I_{\ah}(t,0)$ scales as $(\gamma
v)^{-(D-4)}$. In the ultrarelativistic case $v\simeq 1, \gamma \gg
1$ and combined with $\Lambda$ $\Phi_{\ah}$ scales as

 \begin{align}\label{phi_z_t0_ga_scale}
\Phi_{\ah}(t,0) \simeq - \frac{\vak^2 m   \Gamma(D-4) }{2^{D-1}
\pi^{\fr{D-2}{2}}  \Gamma( \frac{D-2}{2})}     \, \frac{1}{
\gamma^{D-5} |t|^{D-4}}\,\sgn(t)\,.
 \end{align}
Plots of $\Phi_{\ah}$ as function of $r$ with fixed $t>0$ and
ultrarelativistic  $\gamma$ are given in the
Fig.\,\ref{56Dplot_Phiz_r}.

The behavior of curves on these plots near $r=0$ confirms the
$\gamma-$scaling (\ref{phi_z_t0_ga_scale}).

\vspace{0.5em} \textbf{$\boldsymbol{r\gg t, t }$\:fixed.} For
$D=5$ one takes directly (\ref{hyyh}) and (\ref{kotchergah1}) to
deduce:
\begin{align}\label{hyyhff}
  I_{\ah} \simeq \sqrt{\frac{\pi}{2
  } }\frac{1}{r} \xrightarrow[ r\to \infty]{}  0 \,, \quad\quad
 I_{\bh} \simeq \frac{\sqrt{2\pi
  }}{2\,r}\xrightarrow[ r\to \infty]{}  0\,.
\end{align}

\begin{figure}
\begin{center}
\subfigure[\quad  $I_{\ah}$ (black, dashed) and $I_{\bh}$ (red,
solid) in 5D. Both curves possess the same asymptotic behavior
(\ref{hyyhff}).
]{\raisebox{7pt}{\includegraphics[width=7.50cm]{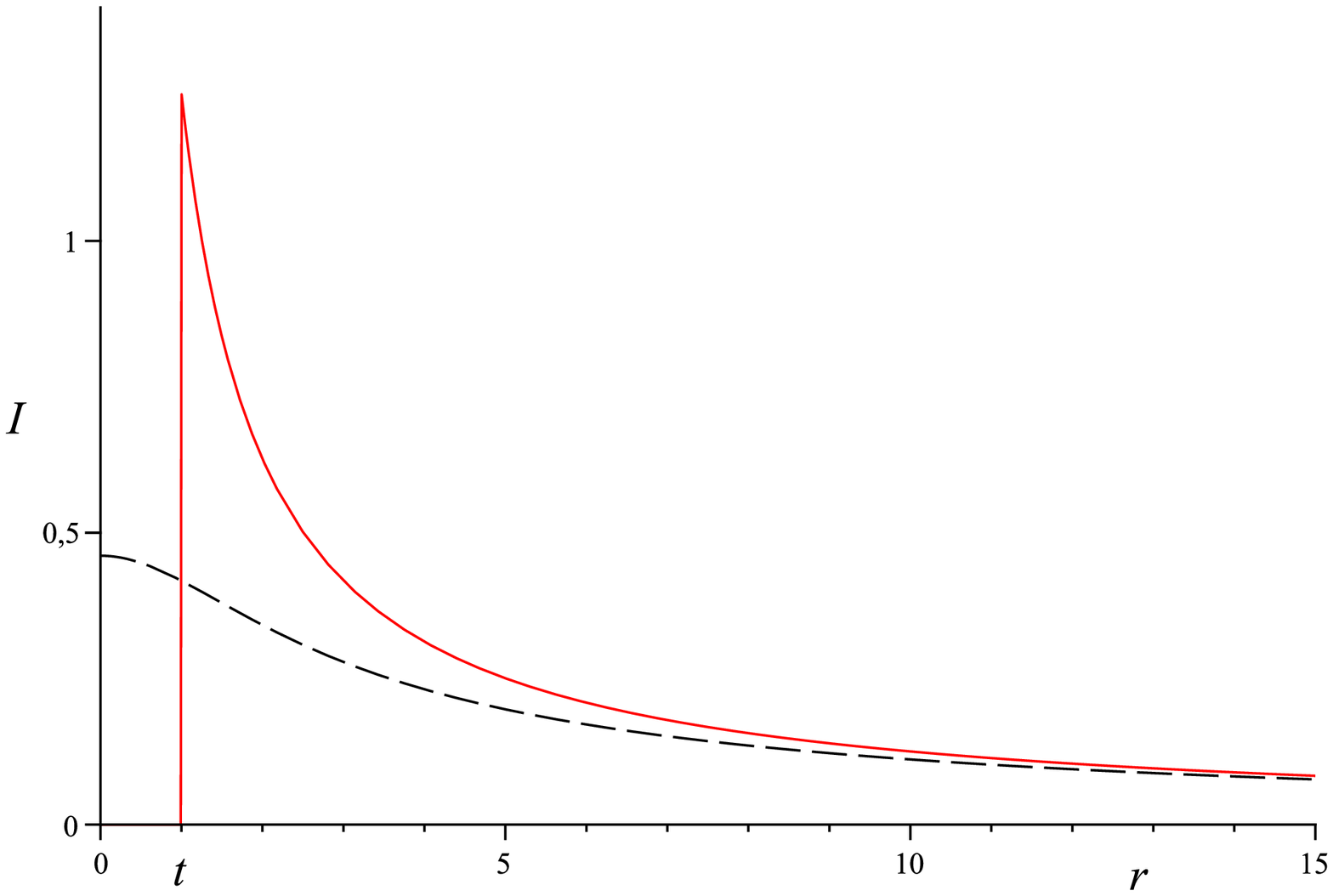}\label{5Dplot_r_I}}}
\hspace{1cm}\subfigure[\quad  $\Phi_{\ah}$ (black, dashdotted),
$\Phi_{\smhsp\bh}$ (green, dashed) and $\Phi$ (red,
solid).]{\includegraphics[width=8.00cm]{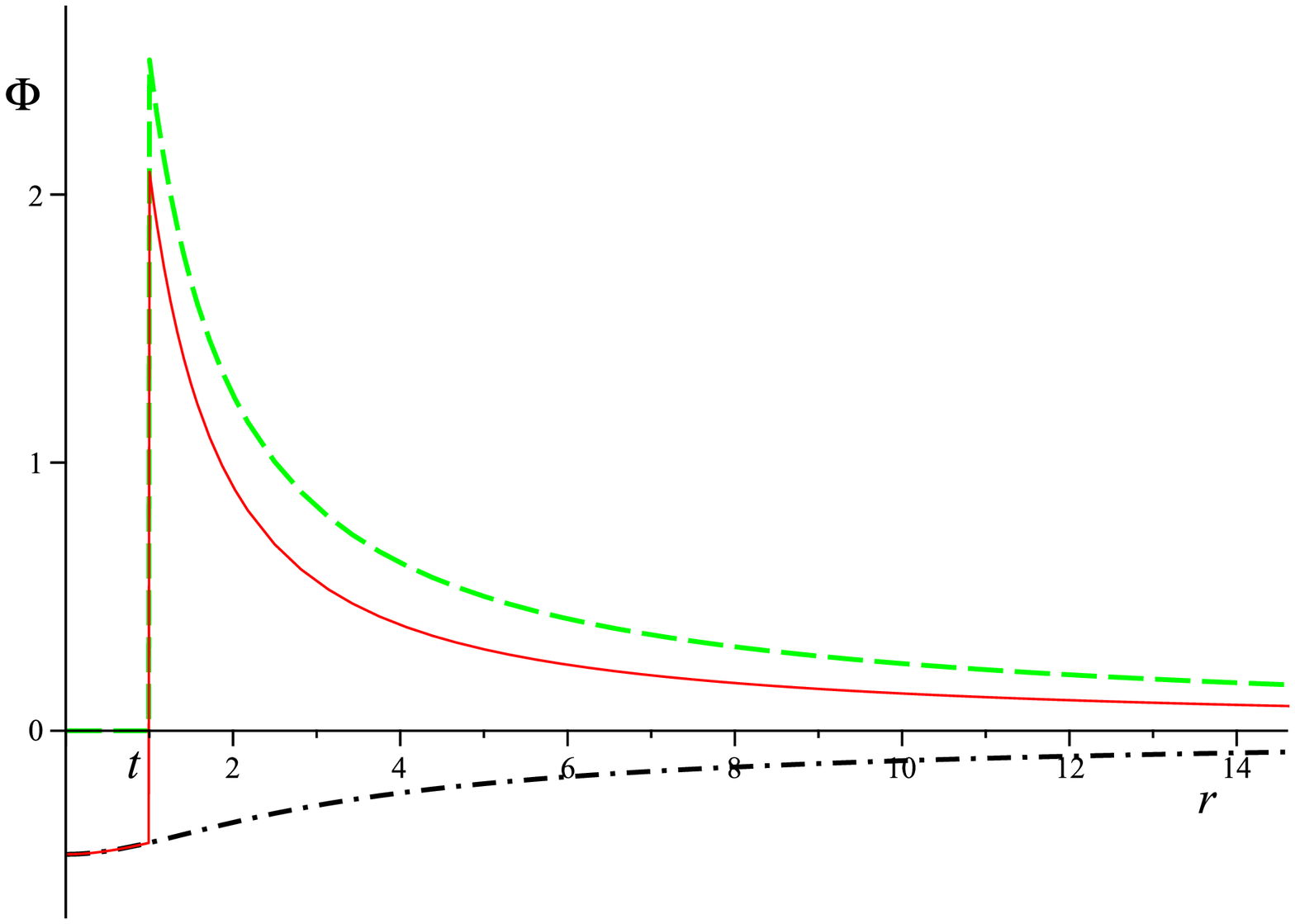}\label{5Dplot_r_Phi}}
\end{center}
\caption{ Dependence of brane perturbations on $r$ for fixed value
of $t$ for $D=5$, $\ga=2$ .} \label{5Dplot_r}
\end{figure}

For $D>5$ we use an asymptotic expansion $\ds J(x) \simeq
\sqrt{\frac{2}{\pi x}}\,\cos(x+\pi/4)$ to estimate
 \begin{align}
 & |I_{\ah}|\leq  \sqrt{\frac{2}{\pi}} \frac{1}{r^{\fr{D-3}{2}}} \int\limits_{0}^{\infty} \! dk
\, {k}^{\frac{D-7}{2}} \, \e^{- k\gamma v|t|}
=\sqrt{\frac{2}{\pi}} \frac{\Gamma\left(\fr{D-5}{2}
\right)}{(\gamma v|t|)^{\fr{D-5}{2}}} \frac{1
}{r^{\fr{D-3}{2}}}\xrightarrow[ r\to \infty]{}  0 \,,
\\  &
|I_{\bh}|\leq \frac{Q(t)}{r^{\fr{D-3}{2}}}\xrightarrow[ r\to
\infty]{} 0, \, \quad\quad Q(t)=\sqrt{\frac{2}{\pi}}
\int\limits_{0}^{\infty} \! dk\,{k}^{\frac{D-7}{2}} \,\cos \, kt\,
.
 \end{align}
Thus both $I_{\bh}$ and $I_{\ah}$ decay at large distances.

\begin{figure}
\begin{center}
\subfigure[ \quad $I_{\ah}$ (black, dashed) and $I_{\bh}$ (red,
solid) in 6D.
]{\raisebox{10pt}{\includegraphics[width=7.50cm]{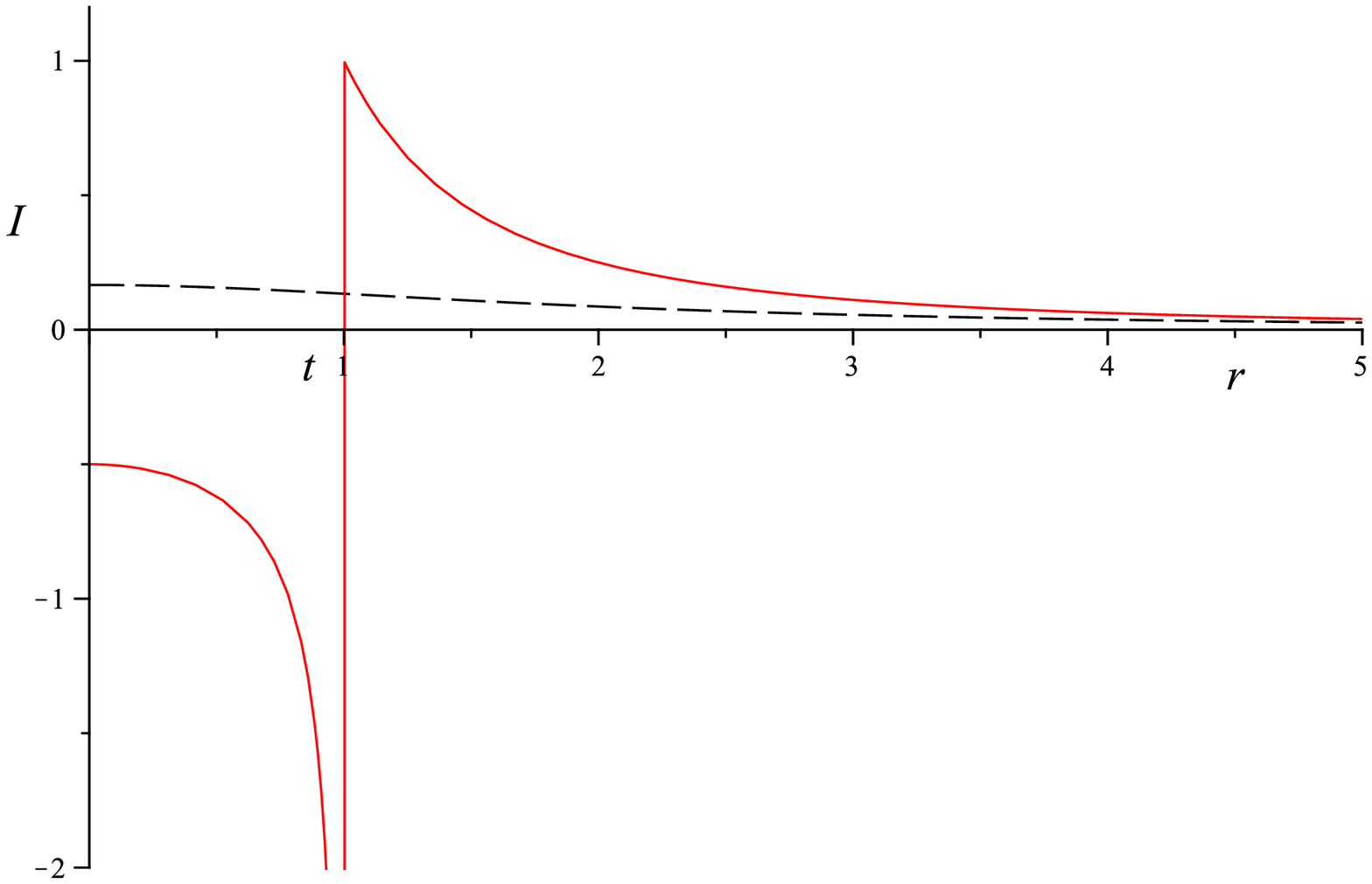}\label{6Dplot_r_I}}}
\hspace{1cm}\subfigure[ \quad $\Phi_{\ah}$ (black, dashdotted),
$\Phi_{\bh}$ (green, dashed) and $\Phi$ (red,
solid).]{\includegraphics[width=8.00cm]{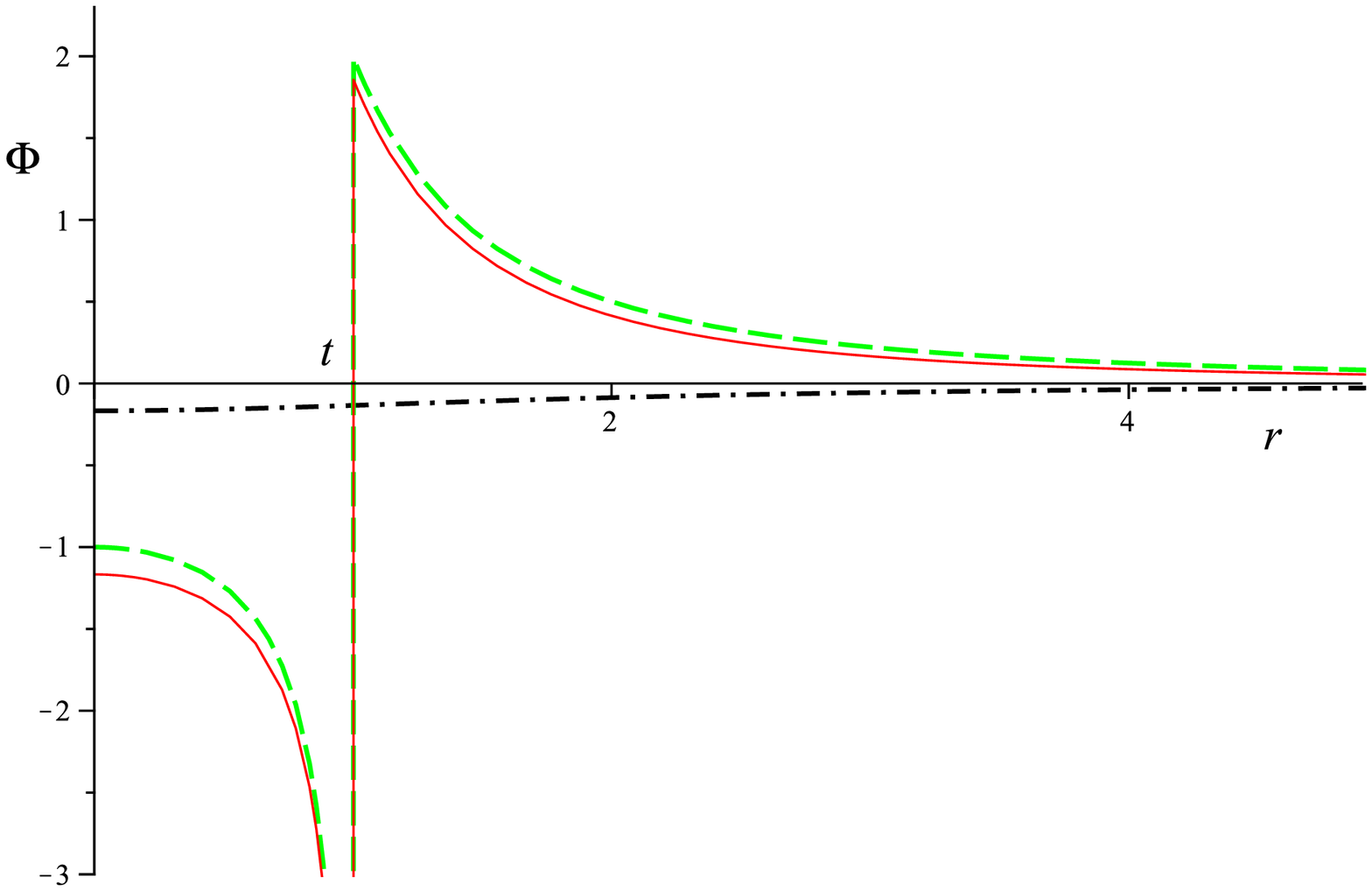}\label{6Dplot_r_Phi}}
\end{center}
\caption{ $D=6$, $\ga=2$ .} \label{6Dplot_r}
\end{figure}

The characteristic behavior of $I$ and $\Phi$ at fixed $t=1$ is
given on Fig.\,\ref{5Dplot_r} ($D=5$) and Fig.\,\ref{6Dplot_r}
($D=6$).

\vspace{0.5em} \textbf{$\boldsymbol{t\gg r,\; r}$\,fixed.} First
consider $D=5$ case: from  (\ref{hyyh}) and (\ref{kotchergah1}) we
deduce directly
\begin{align}\label{hyyhzz}
   I_{\ah}\!\nhsp\left. \vphantom{d^k_K}\right|_{\smhsp
t\gg r}  \simeq \frac{2}{\sqrt{2\pi
  } } \, \frac{1}{\gamma v |t|}   \xrightarrow[ t\to \infty]{}  0\, ,\quad\quad
\lim_{r\to  \infty} I_{\bh}  = \lim_{r \to  \infty}
\frac{\sqrt{2\pi
  }}{2\,r}
\,\theta(r-|t|)=0\,.
\end{align}

\begin{figure}
\begin{center}
\subfigure[ $I_{\ah}$ (black, dashed) and $I_{\bh}$ (red, solid).
]{\raisebox{10pt}{\includegraphics[width=6.50cm]{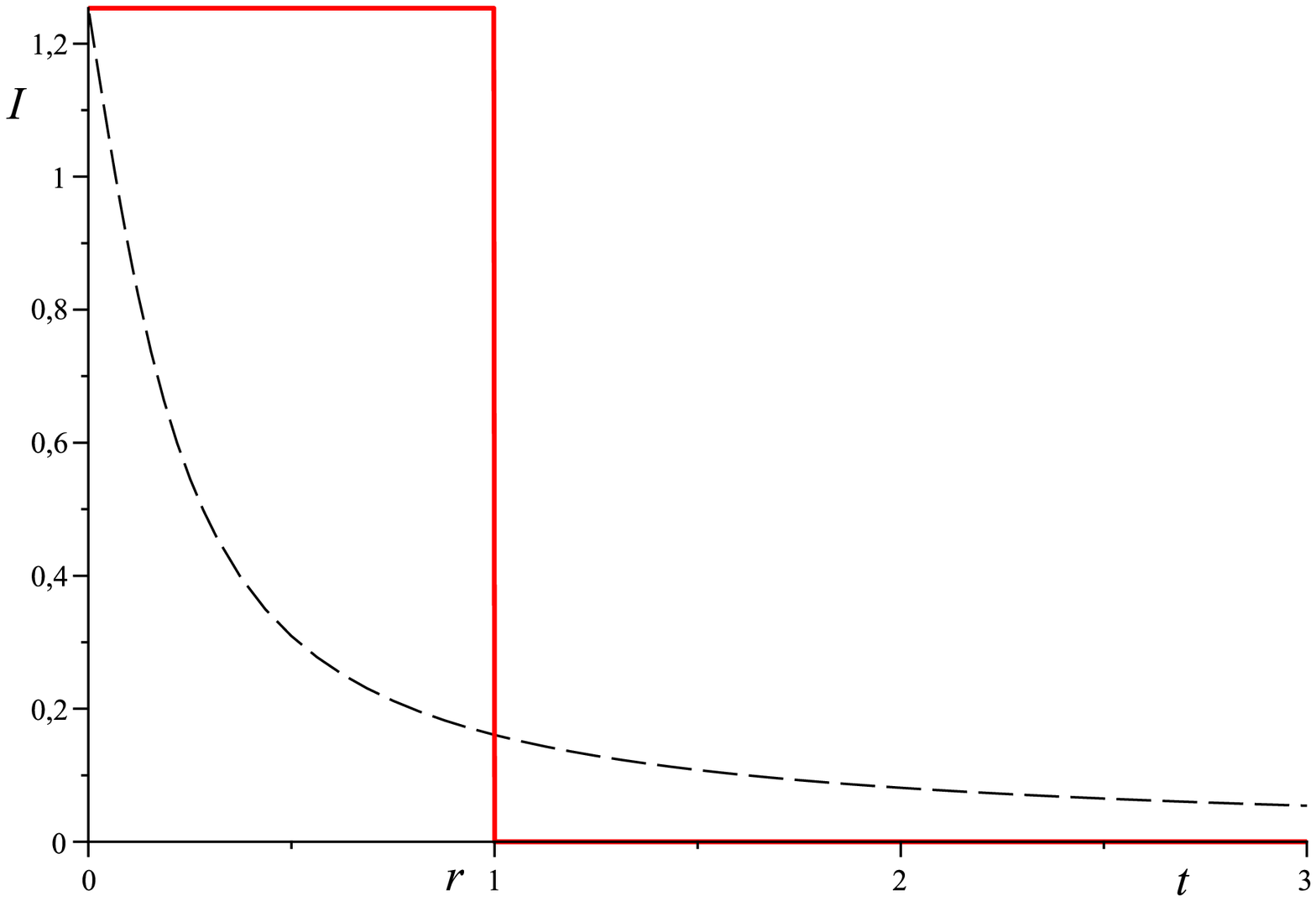}\label{5DplotsI}}}
\hspace{1cm}\subfigure[ $\Phi_{\ah}$ (black, dashdotted),
$\Phi_{\smhsp\bh}$ (green, dashed) and $\Phi$ (red,
solid).]{\includegraphics[width=9.00cm]{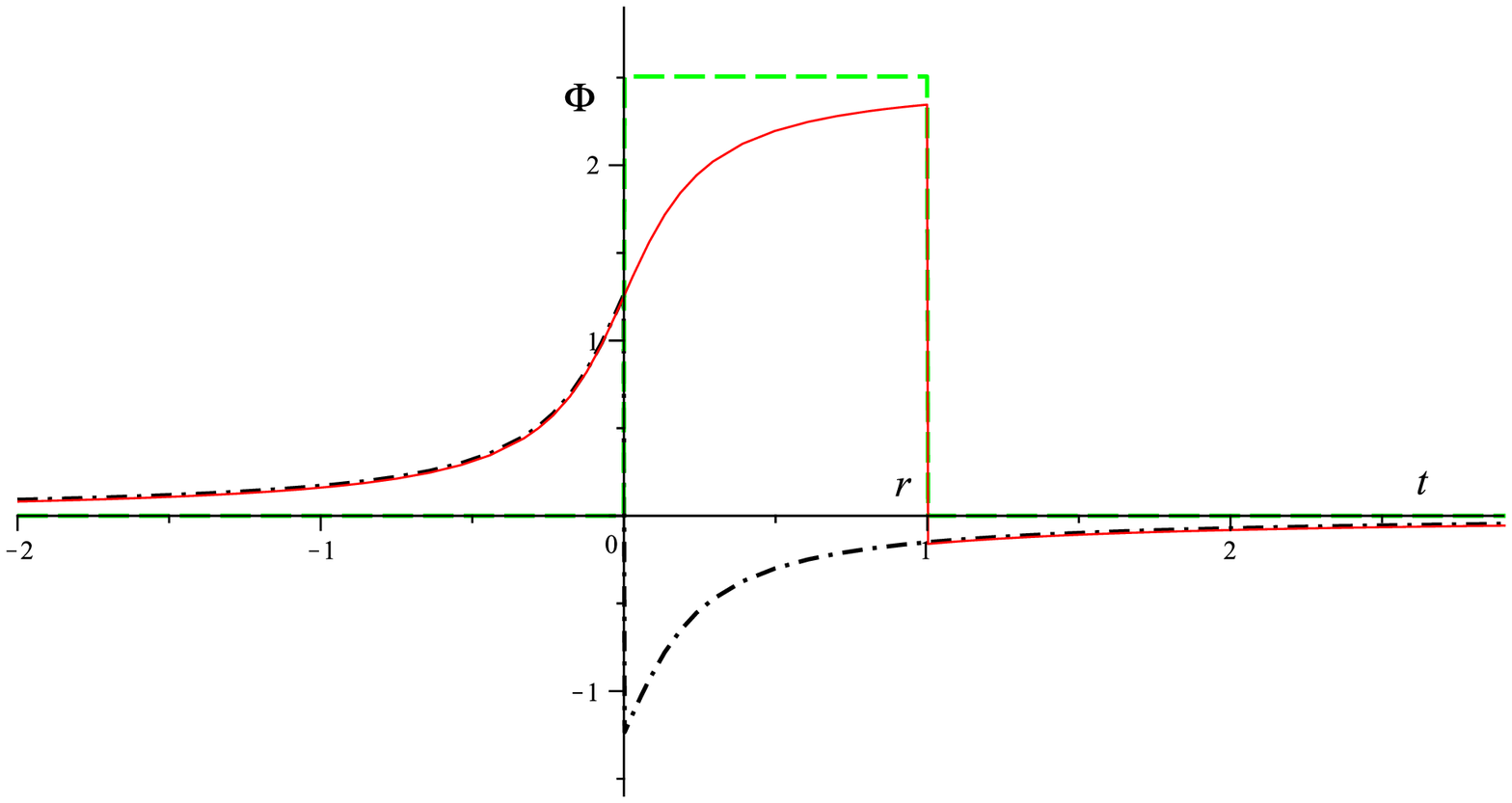}\label{5DplotsPhi}}
\end{center}
\caption{Dependence of brane perturbations on $t$ for fixed value of
$r$ ($D=5$, $\ga=5$) in $r=1$ scale.} \label{5Dplots}
\end{figure}

According to \cite{Watson}, the function
$|x^{-\lambda}J_{\lambda}(x)|$ reaches its absolute maximum at $x=0$
for all $\lambda \geq -1/2$, thus in (\ref{Phi13a}) the Bessel
function is restricted by its  near zero value (\ref{bessel_zero}),
implying
 \begin{align}
 |I_{\ah}| \leq \frac{
\Gamma(D-4)}{2^{\frac{D-4}{2}} \Gamma( \frac{D-2}{2})}  \,
\frac{1}{(\gamma v |t|)^{D-4}} \xrightarrow[ t\to \infty]{} 0
 \end{align}
by virtue of (\ref{sder}).

Now pass to $I_{\bh}$ given by (\ref{Phi13b}): in (\ref{Phi13cn}) we
have shown that the integral
$$
\frac{1}{r^{\fr{D-4}{2}}} \int\limits_{0}^{\infty} \! dk
\,J_{\frac{D-4}{2}}(k r)\,{k}^{\frac{D-6}{2}} $$ converges for all
$D>4$ with the boundary value  (\ref{cic3}). Thus, according to the
Riemann--Lebesgue lemma,
$$\lim_{t\to  \infty} I_{\bh}=0\,.$$

\begin{figure}
\begin{center}
\subfigure[\quad $I_{\ah}$ (black, dashed) and $I_{\bh}$ (red,
solid).
]{\raisebox{0pt}{\includegraphics[width=7.50cm]{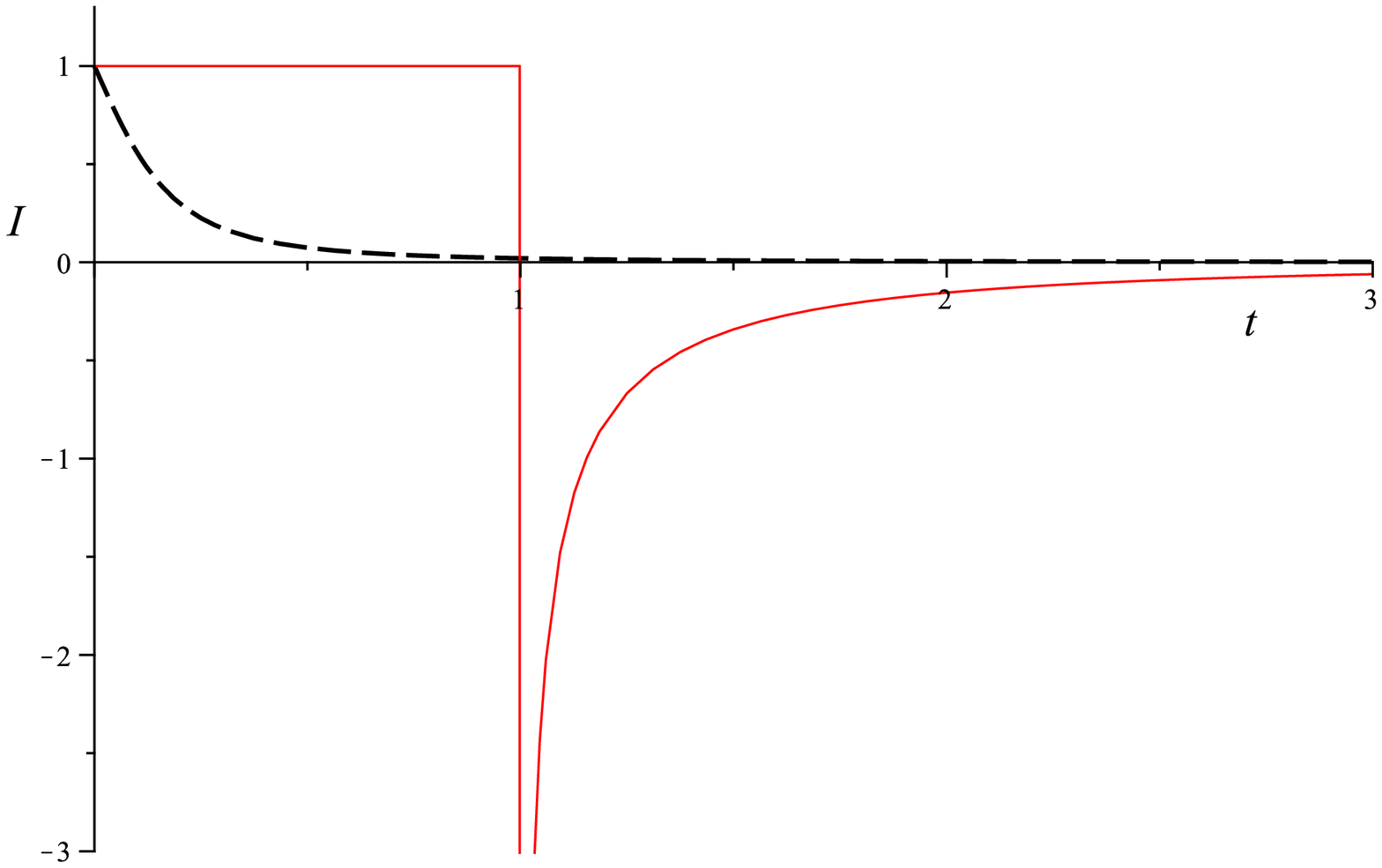}\label{6DplotsI}}}
\hspace{1cm}\subfigure[\quad $\Phi_{\ah}$ (black, dashdotted),
$\Phi_{\smhsp\bh}$ (green, dashed) and $\Phi$ (red,
solid).]{{\includegraphics[width=8.00cm]{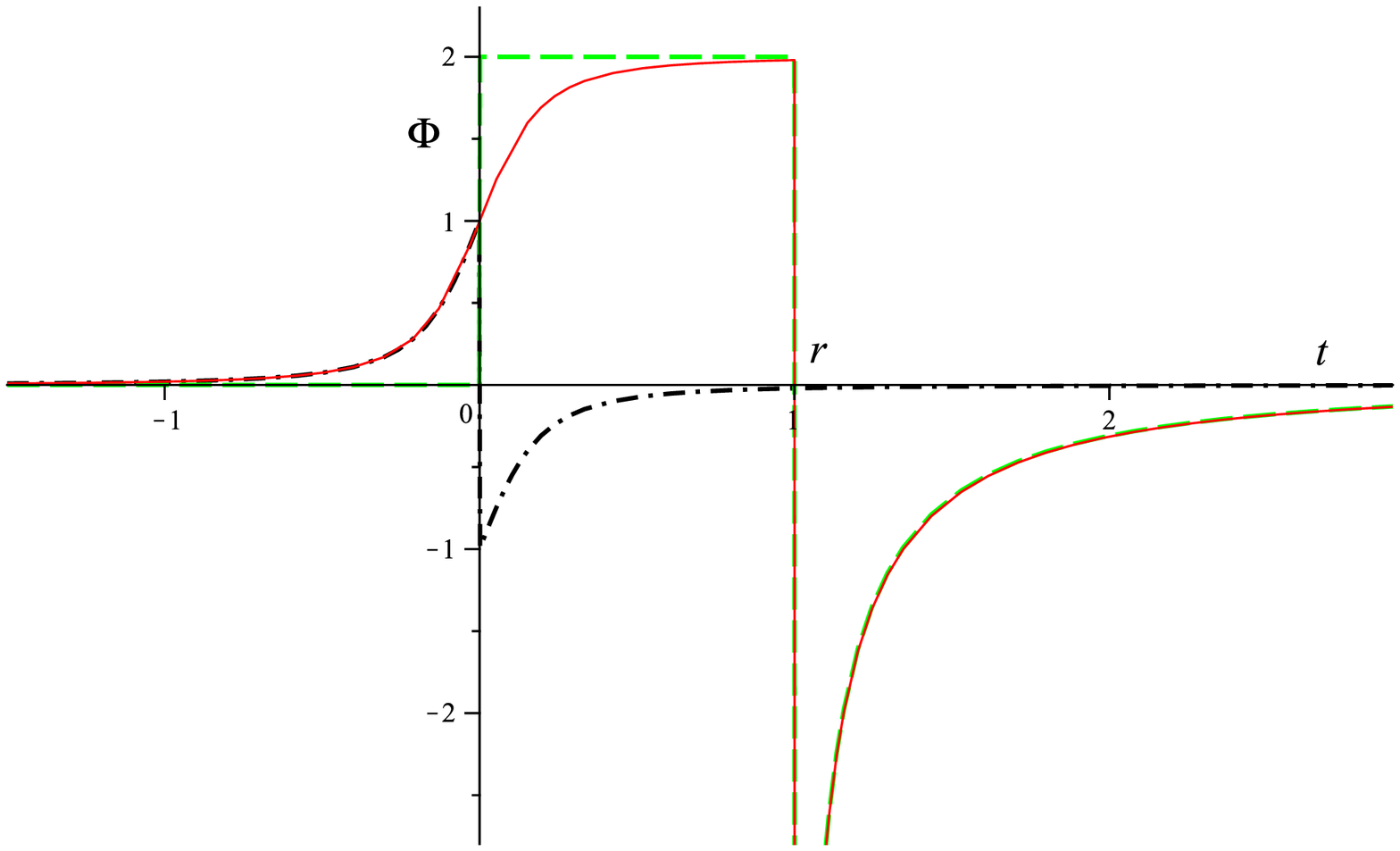}\label{6DplotsPhi}}}
\end{center}
\caption{ $D=6$, $\ga=5$ .} \label{6Dplots}
\end{figure}
The characteristic behavior of $I$ and $\Phi$ at fixed $r=1$ is
given on Fig.\,\ref{5Dplots} ($D=5$) and Fig.\,\ref{6Dplots}
($D=6$).

Note that after the perforation $t>0$  the shock wave
$\Phi_{\smhsp\bh}$ satisfies the homogeneous wave equation,
describing  the free branon wave propagating independently of
further motion of the  particle.   One can notice certain analogy
with the effect of propagation of the string boundary of the hole,
punctured by a body of finite size (such as black hole) in the
domain wall in the field-theoretical models admitting hybrid
topological defects. In that case the string boundary may expand
outwards eating the wall as was argued in \cite{ChamEard}.

\section{Light-like perforation}

It is interesting to consider separately an ultrarelativistic limit
$\gamma\to \infty$, covering also the case of the massless particle.

\subsection{Shock-wave  in arbitrary dimensions}

First we derive the metric generated by a light-like point particle
with the energy $\mathcal{E}$ moving in the $D-$dimensional bulk
along the trajectory $z=t$. It was found that the infinitely-boosted
Schwarzschild metric by Aichelburg and Sexl \cite{Aichelburg} in
four dimensions, being exact, coincides with the solution of
linearized theory with point-like massive source, moving with a
speed of light. For completeness we rederive the linearized solution
in our notation. The field
 (\ref{hpart}), being expressed via the energy $\mathcal{E}$ and the momentum
$p^M$, reads
\begin{align} \label{hpart11}
 \bar{h}_{MN}(x)=-\fr{\vak
\,\mathcal{E}  \Gamma\left(\fr{D-3}{2}\right)}{4\pi^{\fr{D-1}{2}}
 \gamma}
 \left(\frac{\gamma^2}{\mathcal{E}^2}\, p_M p_N-\fr{1}{D-2}\,\eta_{MN}\right)\fr{1}{[\gamma^2(z-v
t)^2+r^2]^{\fr{D-3}{2}}}\,,
 \end{align}
reducing for large $\gamma$ to
\begin{align} \label{hpart12}
 \bar{h}_{MN}(x)=-\fr{\vak
\,   \Gamma\left(\fr{D-3}{2}\right)}{4\pi^{\fr{D-1}{2}}}
 \frac{p_M
 p_N }{\mathcal{E} } \lim_{\gamma \to \infty}
 \fr{\gamma}{[\gamma^2(z-t)^2+r^2]^{\fr{D-3}{2}}}\,.
 \end{align}
Consider first the case $D=5$:
\begin{align} \label{hpart15}
 \bar{h}_{MN}(x)=-\fr{\vak
\, \mathcal{E} }{4\pi^{2}}\, c_M
 c_N \lim_{\gamma \to \infty} \fr{\gamma}{ \gamma^2(z-t)^2+r^2 }\,,
 \end{align}
with $c_M \equiv p_{M} /\mathcal{E}=(1,0,0,0,1) $ being the null
''unit'' vector. Introducing $\alpha=r/\gamma$, one ends up with
\begin{align} \label{hpart17}
 \bar{h}_{MN}(x)=-\fr{\vak
\, \mathcal{E} }{4\pi^{2} r}\, c_M
 c_N \lim_{ \alpha \to + 0} \fr{\alpha}{ (z-t)^2+\alpha^2 }= -\fr{\vak
\, \mathcal{E} }{4\pi r} \,\delta(z-t) \, c_M
 c_N  \,,
 \end{align}
which is just the five-dimensional shock-wave metric.

In  higher dimensions  we start with (\ref{ge_mom}):
\begin{align}\label{ge_mom11}
\bar{h}_{MN}(q) =\frac{2\pi \vak  \, \delta(qp)}{q^2+i \varepsilon
q^0} \left(p_M p_N-\fr{1}{D-2} \frac{\mathcal{E}^2}{\gamma^2}\,
\eta_{MN}\right)\,,
  \end{align}
reducing in the limit $v=1$, $\gamma \to \infty$ to:
\begin{align}\label{ge_mom12}
   \bar{h}_{MN}(q) =\frac{2\pi \vak  \, \mathcal{E} \, \delta(q^0-q^z)}{q^2+i \varepsilon q^0}  \, c_M c_N \,.
  \end{align}
Inverting the Fourier-transform according to the definition
(\ref{fur}) and integrating over $q^0$ we obtain
\begin{align}\label{ge_mom13}
\bar{h}_{MN}(x) =- \frac{ \vak  \, \mathcal{E}}{(2\pi)^{D-1}}  \,
c_M c_N \int \frac{  \e^{ i \mathbf{q} \mathbf{r}}}{\mathbf{q}^2}
\, d^{D-2} \mathbf{q} \int \e^{ i q^z(z-t)}  d q^z \,.
  \end{align}
This gives the product of $\delta(z-t) $ with the Coulomb potential
in $\mathbb{R}^{D-2}$, so in $D>4$ one has:
\begin{align}\label{ge_mom14}
h_{MN}(x) =- \frac{ \vak  \, \mathcal{E}
\Gamma\left(\fr{D-4}{2}\right)\, \delta(z-t) }{4\, \pi
^{\frac{D-2}{2}}r^{D-4}}  \, c_M c_N    \,
  \end{align}
 and in four dimensions
\begin{align}\label{ge_mom15}
\bar{h}_{MN}(x) =  \frac{ \vak  \, \mathcal{E} \, \delta(z-t)
}{2\, \pi} \, \ln r \,c_M c_N    \,.
  \end{align}
 For $D=5$ (\ref{ge_mom14}) coincides with (\ref{hpart17}).

Going back from (\ref{ge_mom14}) to  (\ref{hpart12}) one finds the
following identity for the delta-like sequence:
\begin{align}\label{ident}
\lim_{ \alpha \to + 0} \fr{\alpha^{2n-1}}{{\rule{0cm}{0.8em}[
x^2+\alpha^2]}^{n} }=\frac{\sqrt{\pi }\,
\Gamma\left(n-\fr{1}{2}\right)}{\Gamma(n)}\, \delta(x)\,.
  \end{align}

\subsection{Branon solution}
The source term in the branon equation in the light-like limit
reads:
 \begin{align}\label{jxb1}
\lim_{\gamma\to \infty} J(x)&=-\fr{\vak^2
 \mathcal{E}  \Gamma\left(\fr{D-1}{2} \right)}{4\pi^{\fr{D-1}{2}}}\fr{
\gamma^3t}{(\gamma^2 t^2+r^2)^\fr{D-1}{2}}\,= \non\\&
=\fr12\fr{\vak^2 \mathcal{E}  \Gamma\left(\fr{D-3}{2}
\right)}{4\pi^{\fr{D-1}{2}}}\frac{\pa}{\pa t} \fr{\gamma}{[\gamma^2
t^2+r^2]^\frac{D-3}{2}} = \fr{\vak^2 \mathcal{E}
\Gamma\left(\fr{D-4}{2}
\right)}{8\pi^{\fr{D-2}{2}}r^{D-4}}\,\delta'(t)\,.
 \end{align}
 Thus the source flashes only at the moment of
perforation $t=0$, and the domain wall gets excited only after the
collision. The surviving term in (\ref{jxb1}) should serve as a
source for both the antisymmetric part $\Phi_{\ah}$   and the branon
component $\Phi_{\smhsp\bh}$ of the full deformation. However, one
can see that the antisymmetric part is absent for the light-like
perforation. Indeed, for $D=6$ the limit of (\ref{hyyhrffr}) is zero
\begin{align}\label{hyyhrffr_UR}
I_{\ah} =\frac{1}{r^{2}} \lim_{\gamma \to
\infty}\left[1-\frac{\gamma v |t|}{ \sqrt{\gamma^2 v^2
t^2+r^2}}\right]=\lim_{\gamma \to \infty}\left[\frac{1}{2(\gamma v
t)^2}+\mathcal{O}\hsp(\gamma^{-4})\right]=0\,.
\end{align}
In higher even dimensions $I_{\ah}$ can be obtained by consecutive
application of the operator $-2\partial/\partial r^2$.  In accord
with the Leibnitz rule,  differentiation does not affect the
$\gamma\to \infty$-limit, the resulting expression
$$ \left(- \frac{1}{r} \frac{\partial}{\partial r}
\right)^{\!m}\left[1-\frac{\gamma v |t|}{ \sqrt{\gamma^2 v^2
t^2+r^2}}\right] =\frac{(2m-1)\,\gamma v |t|}{ (\gamma^2 v^2
t^2+r^2)^{m+1/2}}\,, \quad\quad m\geq 1
$$ going to zero as $\gamma\to
\infty$ too.

In odd dimensions   (\ref{hyyh}) the leading in $\gamma$
contribution  to $ I_{\ah} $ is given by
\begin{align}\label{hyyh_lim}
  I_{\ah} = \frac{2}{\sqrt{2\pi
  }  }  \left(- \frac{1}{r} \frac{\partial}{\partial r}
\right)^{\!m}  \frac{1}{r}\,\arctan \frac{r}{\gamma v |t|} \simeq
\frac{2}{\sqrt{2\pi
  }}\,
 \frac{(2m-1)!! \, (\gamma v |t| )^{2m}}{r^{2m+1}(\gamma^2 v^2
t^2+r^2)^m} \,\arctan \frac{r}{\gamma v |t|} \, , \quad\quad m\geq 0
\end{align}
and it is also vanishing in the limit $\gamma\to \infty$.

Next, coming to $\Phi_{\smhsp\bh}-$part, we see that, taking into
account (\ref{hyyhrffr_UR}), the Eq. (\ref{jajcc2})  tends to
 \begin{align}\label{jajcc2_UR}
\lim_{\gamma \to \infty} \Box_{D-1}{\Phi}_{\smhsp\bh}=
\frac{\vak^2 \Eps
     \Gamma \left(
 \fr{D-4}{2}  \right) }{ 8 \pi^{\fr{D-2}{2}}
r^{ D-4}  }\, \delta'(t)\,
   =\lim_{\gamma \to
\infty} \Box_{D-1}{\Phi}\, ,
\end{align}
in agreement with (\ref{jxb1}). We conclude that in the light-like
limit the full deformation is  ${\Phi}={\Phi}_{\smhsp\bh}$, where
 \begin{align}\label{Phi_w_UR}
 \Phi_{\smhsp\bh} =  \frac{\vak^2 \Eps}{(2\pi )^{{D}/{2}}  } \,
\theta(t) \,I_{\bh}\,,
\end{align}
with $I_{\bh}$ given by (\ref{hyyhw2}) and (\ref{jaj3}) for odd and
even bulk dimensions, respectively.

\section{Perforation of the domain wall in four dimensions}

The case $D=4$ requires special consideration since some
intermediate formulas fail. It is also physically most interesting
allowing for applications to the cosmological domain wall problem.
In fact, the primordial domain walls moving in the ultrarelativistic
gas should be excited via the mechanism described in the previous
section.

The linearized fields generated by the particle and the domain wall
will read  in $D=4$:
 \begin{align}\label{4Dpart}
   &
 \bar{h}_{MN}(x)=-\fr{\varkappa_4
\hsp m}{4\pi}
 \fr{ u_M
u_N-\fr{1}{2}\,\eta_{MN}}{\rule{0mm}{1em}[\gamma^2(z-v
t)^2+r^2]^{1/2}}\,,
 \\ \label{4Dwall}
   &h_{MN}(x)=\frac{\varkappa_4\hsp \mu
|z|}{4}\,{\rm diag}(-1,1,1,{3})\, .
\end{align}
  The branon equation (\ref{NGEQ}) takes the form
\begin{align}\label{breq_4D}
\Box_3 \Phi= - \fr{\la vt}{[\gamma^2 v^2 t^2+r^2]^{3/2}}\,,\quad
\quad \la=\fr{\varkappa_4^2 m\gamma^2  }{8\pi}\left( \gamma^2v^2
+\fr{1}{2}\right)\,.
\end{align}
Trying naively to adapt for $D=4$ the integral (\ref{Phi11})
 \begin{align}\label{Phi11_4D}
\Phi=    \frac{\la}{ \gamma^3 } \int\limits_{0}^{\infty} \! dk
\,\frac{J_{0}(k r)}{k} \left(-\sgn(t)\, \e^{- k\gamma v|t|}+
2\,\theta(t)\,\cos \, kt
  \right)\, ,
 \end{align}
one finds that it diverges at $k=0$, so we need some regularization.
To achieve this goal we notice that the solution  in the next even
dimension $D=6$ must follow from the $D=4$ solution  by
differentiation rules described in Sec. 5.  Also taking into account
that splitting of the full $\Phi$ into two parts for $D>4$ is
related to causality which is preserved by the differentiation over
$r$, we can set following requirements on the $D=4$ solution:
\begin{itemize}
    \item the action of $\ds -\frac{1}{r}\frac{\partial}{\partial
    r}$ must generate separately the six-dimensional solutions (\ref{jaj1}) and
    (\ref{hyyhrffr}) ;
        \item the higher-dimensional expressions remaining regular
        in the limit $D\to 4$  should be trusted in the $D=4$ case,
        namely

        -- the static part $I_{\ah}^{\rm stat}$ should satisfy
         the condition
    (\ref{les_freles_roseaux_2}), while  the dynamical part
    $I_{\ah}^{\rm dyn}$ have to coincide with (\ref{hypergeom_dyn});

-- the matching conditions should hold: $ I_{\ah}\!\nhsp\left.
\vphantom{d^k_K}\right|_{\smhsp t=0}=I_{\bh}\!\nhsp\left.
\vphantom{d^k_K}\right|_{\smhsp t=0} $;

-- $\Box \Phi_{\ah}$ has to agree  with (\ref{cic2}) and $\Box
\Phi_{\smhsp\bh}$ has to agree with (\ref{jajcc3_mod});

\item both $I_{\ah}$ and $I_{\bh}$ have to be the even functions
of $t$ while $\Phi_{\ah}$ and $\Phi_{\smhsp\bh}$ should be defined
via $I_{\ah}$ and $I_{\bh}$ in the same way as in
(\ref{Phi13a},\ref{Phi13b}).

\end{itemize}

Starting with $I_{\ah}^{\rm stat}$, we see that the Eq.
(\ref{les_freles_roseaux_2}) is regular at $D=4$, so using the
relevant Green's function we get
 \begin{align}\label{Mizulina_merde}
\Delta_3 I_{\ah}^{\rm  stat} =-2\pi\,\delta^3(\mathbf{r})\,,
\quad\quad I_{\ah}^{\rm stat}=-\ln r\,.
 \end{align}
Next, the  dynamical component $I_{\ah}^{\rm dyn}$ is given by the
hypergeometric function in (\ref{hypergeom_dyn}):
 \begin{align}\label{Mizulina_merde_2}
I_{\ah}^{\rm dyn} = -   \frac{ \gamma v |t| }{r  } \;  {}_2
{}F_1\! \left(  \frac{1}{2}\,, \frac{1}{2} \,;
 \frac{3}{2}\,; -\frac{\gamma^2 v^2 t^2}{r^2}\right)
\frac{ }{
 } =-  \mathrm{Arsh}  \frac{\gamma v |t|}{r} \,,
\end{align}
which is analytical for any values of $t$ and $r$. In order to check
the differentiation rule it is convenient to modify the inverse
trigonometric functions as\hsp
 \footnote{$I_{\ah} $ presented here, does not satisfy the
condition $I_{\ah} \to 0$ for $\gamma \to \infty$ only \textit{with
logarithmic precision}: $ \ln (\frac{r}{2 \gamma v |t|})+
\mathrm{Arth} \frac{\gamma v |t|}{\sqrt{\gamma^2 v^2 t^2+r^2}}\to
0$. Notice that this condition is not  mandatory.}:
\begin{align}\label{hyyhrffr_4D}
I_{\ah} = -\ln r- \mathrm{Arth} \frac{\gamma v |t|}{\sqrt{\gamma^2
v^2 t^2+r^2}} \,.
\end{align}
Now one can easily see that (\ref{hyyhrffr_4D}) does coincide with
(\ref{hyyhrffr}) upon action of $2\,\partial/d r^2$, with obvious
correspondence to static and dynamical parts of (\ref{hyyhrffr}).
Thus $\Phi_{\ah}$  becomes
\begin{align}\label{hypergeom_4D}
\Phi_{\ah} = \frac{
 \la }{ \gamma^3}\, \left(   \mathrm{Arsh}  \frac{\gamma v t}{r}
   +   \sgn(t)
 \ln
r \right)\,,
\end{align}
with the jump at $t=0$ equal to
 \begin{align}\label{jump_univ_4D}
 \delta  \Phi_{\ah}  =  \frac{
 2\,\la   }{ \gamma^3} \, \ln r  \,.
\end{align}
One can check that
 \begin{align}\label{jajcc3_4D}
\Box_{3}{I}_{\ah}=  \frac{ \gamma^3 v|t|}{[\gamma^2 v^2
t^2+r^2]^{3/2}}\, , \quad \quad \Box_{3}{ \Phi}_{\ah} =  \frac{
2\smhsp \la  \, \ln r }{ \gamma^3
   }\, \delta'(t)-\frac{\la vt}{[\gamma^2 v^2
t^2+r^2]^{3/2}}\,    ,
\end{align}
in agreement with  (\ref{cic2}) and (\ref{jajcc3_mod})
\footnote{Making use of the identity $\mathrm{Arsh}\hsp
x=\ln\!\left(x+\sqrt{x^2+1}\right)$, $I_{\ah}$ (\ref{hyyhrffr_4D})
can be equivalently presented as
$$I_{\ah}=-\ln\!\left(\gamma v |t|+\sqrt{\gamma^2 v^2 t^2+r^2}
\right)\,.$$ Acting by the box on $I_{\ah}$ we do not get the
$\delta^2(\mathbf{r})$-term.}.

Finally, turning  to the shock-wave part $I_{\bh}$, we see that the
action of $2\smhsp\partial/\partial r^2$ must give (\ref{jaj1}). The
expression (\ref{jaj1}) contains two step functions $\theta(r-t)$
and $\theta(t-r)$, projecting the continuous functions onto the
non-overlapping domains before and after the wave-front $t=r$.
Therefore, both components of (\ref{jaj1}) should be obtainable by
the action of $2\smhsp\partial/\partial r^2$ on the corresponding
terms \emph{separately}, without generating terms localized on the
front $t=r$. In other words, $I_{\bh}$ should be continuous. Thus,
multiplying (\ref{jaj1}) by $r$ and integrating, one obtains:
 \begin{align}\label{jaj1_4D}
I_{\bh}= -\ln r\: \theta(r-|t|) - \ln\nhsp
\left(|t|+\sqrt{t^2-r^2}\right) \,\theta(|t|-r) \,.
\end{align}
Note that the  derivatives of $I_{\bh}$ over $r$ do not contain
$\delta(|t|-r)$ (canceled between two terms), so the continuity
requirement is preserved and (\ref{jaj1}) is reproduced under the
action of $2\smhsp\partial/\partial r^2$ indeed.

It remains to check that the expression (\ref{jaj1_4D}) does satisfy
the homogeneous d'Alembert equation. Acting on it by the box we
obtain
 \begin{align}\label{moi0}
\Box \left[\hsp \ln\nhsp
\left(|t|+\sqrt{t^2-r^2}\right)\right]=\frac{2+\ln
|t|}{r}\,\delta(|t|-r)=-\Box \left[\vp \ln r\:
\theta(r-|t|)\right],
\end{align}
so $\Box I_{\bh}=0$ indeed.

Hence, according to  (\ref{jajcc2}), $\Box \Phi_{\smhsp\bh}$ is
determined by the value $I_{\bh}\!\nhsp\left.
\vphantom{d^k_K}\right|_{\smhsp t=0}=- \ln r$. It is equal to
$I_{\ah}\!\nhsp\left. \vphantom{d^k_K}\right|_{\smhsp t=0}$ and
satisfies  the condition (\ref{hyyhrffr_4D}). The  required jump
condition $\delta \Phi_{\smhsp\bh}=-\delta\Phi_{\ah}$ is  satisfied
too by virtue of relations between $\Phi_{\ah}$  and
$\Phi_{\smhsp\bh}$ with $I_{\ah}$ and $I_{\bh}$, respectively.

Thereby   $\Phi_{\smhsp\bh}$ satisfies
 \begin{align}\label{jajcc2_4D}
&\Phi_{\smhsp\bh}= -2 \Lambda \,\theta(t) \left[\hsp \ln r\:
\theta(r-t)+\ln\nhsp \left(t+\sqrt{t^2-r^2}\right)
\,\theta(t-r)\right] \,, \\ &\Box_{3}{\Phi}_{\smhsp\bh}= -\frac{
 2\smhsp \la  \hsp\ln r }{  \gamma^3  }\, \delta'(t)\,
,
\end{align}
in agreement with (\ref{jajcc2}).

\section{Summary and concluding remarks}
In this paper we have considered the piercing collision of a point
particle with the Nambu-Goto domain wall within the linearized
gravity. We have shown that, in spite of the singular nature of the
particle's gravitational field at the perforation point in
space-time, this process can be safely described in terms of
distributions. From our analysis it follows that the domain wall
gets excited after the perforation via creation of the spherical
branon shock wave propagating outwards with the speed of light. This
wave is the reaction of the wall on the instantaneous change of
particle's acceleration on a finite amount due to homogeneous nature
of the domain wall gravitational field.

This effect completes the standard picture of interaction of domain
walls with particles  developed earlier to describe propagation of
the domain walls through matter in the early universe \cite{Vilsh}.
At low temperature most of the surrounding particles are reflected
from the wall transferring to it some momentum. This results in the
friction force acting on the domain wall. We have shown that those
particles, which are {\em not reflected}, do not transfer momentum
to the wall, but cause an excitation effect. Since we have
restricted our consideration here by the linearized approximation,
the resulting energy dissipation could not be seen, but  in   fact
it can be expected in the form of gravitational radiation once
higher-order effects are included. For this one has to extend the
theory to the next post-linear order similarly to the case of
colliding particles \cite{Galtsov:1980ap,GKST2,GKST3,GKST4}.

One could notice certain analogy between our effect and an
essentially non-linear phenomenon which may accompany perforation of
the field-theoretical domain wall in the models admitting hybrid
wall/cosmic string defects. Namely, the hole in the wall may be
surrounded by a cosmic string which either collapses restoring the
initial shape of the domain wall  or expands to infinity completely
``eating'' the  wall \cite{ChamEard}. This was suggested as
mechanism of  destruction of domain walls in the early universe.
Further investigation have shown  that the issue of the above
dilemma depends on the number of the created holes: for destruction
to occur it is necessary that at least four such holes were produced
in the domain wall \cite{Stojkovic:2005zh}. Of course these effects
can not be touched within our approach, but  their nature is still
 similar : both are related to an intrinsic tension of the
relativistic extended objects. Moreover, it is likely that branon
excitations in the non-linear field-theoretical domain wall --
black hole system were already observed in numerical experiments
\cite{Flachi:2007ev}.

As another line of applications, we could mention the
Rundall-Sundrum type scenarios \cite{RS, RS1, RS2}. Bulk matter
perforating the brane representing our universe could generate
``unmotivated'' explosive events \cite{GMZ} in the observed
Universe.  Indeed, the branon field universally interacts with
matter on the brane via the induced metric \cite{KuYo,Bu, Bu2},
and, in the non-linear regime the branon explosion will transform
into the matter explosion. Speculating further, one could mention
that  in the brane-world scenarios the branons appear as  new
relevant low-energy particles \cite{Sundrum:1998sj,Dobado:2000gr}
whose couplings are suppressed by the brane tension
\cite{Bando:1999di}. They are therefore generically stable and
weakly interacting, the features making them natural dark matter
candidates \cite{Cembranos:2003fu} (for a recent review see
\cite{Cembranos:2013qja}). Our mechanism of branon excitation by
bulk matter should be explored in this context as well.

\section*{Acknowledgments}
This work was supported by the RFBR grant 11-02-01371-a. PS is
grateful to non-commercial ''Dynasty'' foundation (Russian
Federation) for financial support.

\appendix

\section{Computation of the Fourier-transforms}\label{app2}
Here we give a typical calculation of  the Fourier-transforms used
in this paper. Consider the branon source in the coordinate space
 $J( \sigma)$ as given by (\ref{jxb}):
 \begin{align} \label{jxbff}
 J( \sigma )=  - \fr{\la vt}{[\gamma^2 v^2
t^2+r^2]^\fr{D-1}{2}}\, .
 \end{align}
Substituting this to  (\ref{fur}) one  introduces spherical angles
such that $d^{D-2} r =\Omega_{D-4} r^{D-3} \sin^{D-4}\theta \,dr \,
d\theta$ with $\theta$ being the angle to the polar axis chosen
along $\mathbf{k}$:
\begin{align}
J(k)= -\lambda v \Omega_{D-4} \int   \fr{t\, e^{i  (\omega t -kr
\cos \theta)}}{[\gamma^2 v^2 t^2+r^2]^\fr{D-1}{2}}  \,
r^{D-3}\sin^{D-4}\theta\,  dt\,
 \,dr \, d\theta .
 \end{align}
Combining the  integral \cite{GR}
\begin{align}
 \int\limits_{0}^{\pi} e^{\pm i z \cos \theta}\sin^{2k}\!\theta\, d\theta =
 \frac{2^{k}\Gamma(k+1/2)\sqrt{\pi}}{z^k}\, J_k(z)
\end{align}
 with the volume measure  $\Omega_{n-1}$,
the full angular integral of $\e^{\pm i z \cos \theta}$ can be
presented as
 \begin{align}\label{totangint}
\int \e^{\pm i z\cos \theta}d\Omega_{n}=\fr{(2\pi)^{
\frac{n+1}{2}}}{ {z}^\fr{n-1}{2}} J_{\fr{n-1}{2}}(z)\,.
 \end{align}
Thereby one gets
\begin{align}
J(k)= -\frac{(2\pi)^{ \frac{ D-2}{2}} \lambda v}{{ k
}^{\frac{D-4}{2}}}   \int \fr{t\, e^{i  \omega t  }
r^{\frac{D-2}{2}}}{[\gamma^2 v^2 t^2+r^2]^\fr{D-1}{2}}  \,
J_{\frac{D-4}{2}}(kr)  \, dt\,
 \,dr   .
 \end{align}
Integration over $r$ is done using the integral:\cite{Proudn}
\begin{align}\label{baraboulka0}
 \int\limits_{0}^{\infty} dy \frac{y^{n+1}J_n(b y)}{[y^2+a^2]^{m}}=\frac{b^{m-1} |a|^{n-m+1}}{2^{m-1}\Gamma(m)}\, K_{n-m+1}
 (b\hsp
 |a|)\, ,
\end{align}
and then
\begin{align}
J(k)= -\frac{ \pi^{ \frac{ D-1}{2}} \lambda  }{\Gamma\left(\frac{
D-1}{2}\right) \gamma }   \int\limits_{-\infty}^{\infty}  \e^{i
\omega t } \e^{-\gamma v k |t|} \,{\rm sgn}\, t\, dt\,  ,
 \end{align}
where only the imaginary part of the exponent survives by parity.
Finally, one ends up with
\begin{align}\label{Jk}
J(k)= -\frac{2  \pi^{ \frac{ D-1}{2}} i \lambda
}{\Gamma\left(\frac{ D-1}{2}\right) \gamma }
\int\limits_{0}^{\infty}  \e^{-\gamma v k  t } \sin { \omega t }
\, dt =- \frac{2 \pi^{ \frac{ D-1}{2}}  \lambda
}{\Gamma\left(\frac{ D-1}{2}\right) \gamma }
\frac{i\omega}{\omega^{2}+\gamma^{2} v^{2} k^{2}}\, .
 \end{align}

In the context of the \RS models we need to require that the
dominant contribution in the $r-$integration will come from the
region of validity of linearized approximation, which means
$$r<1/k\,,\qquad \gamma v t <1/k\,. $$

\end{document}